\pdfoutput=1
\documentclass[11pt,a4paper]{article}
\usepackage{jheppub,amsfonts,mathrsfs}
\usepackage[english]{babel}
\usepackage[caption=false]{subfig}

\newcommand{\beq}{\begin{eqnarray}}
\newcommand{\eeq}{\end{eqnarray}}
\newcommand{\non}{\nonumber\\}
\newcommand{\p}{\partial}
\DeclareMathOperator{\Tr}{Tr}
\DeclareMathOperator{\tr}{tr}

\newcommand{\dphi}{\delta{\mkern-1mu}\phi}

\newcommand{\dA}{\delta{\mkern-2mu}A}
\newcommand{\da}{\delta{\mkern-1mu}a}
\newcommand{\dH}{\delta{\mkern-1mu}H}

\DeclareMathOperator{\U}{U}
\DeclareMathOperator{\SU}{SU}
\DeclareMathOperator{\SO}{SO}
\DeclareMathOperator{\USp}{USp}

\DeclareMathOperator{\GL}{GL}
\DeclareMathOperator{\SL}{SL}

\renewcommand{\i}{\mathrm{i}}
\renewcommand{\d}{\mathop{}\!\mathrm{d}}

\newcommand{\calI}{\mathcal{I}}
\newcommand{\bbD}{\mathbb{D}}
\newcommand{\NF}{N_{\rm f}}

\newcommand{\NC}{N}

\DeclareMathOperator{\kernel}{kernel}
\newcommand{\sfM}{{\sf M}}

\newtheorem{theorem}{Theorem}
\newtheorem{corollary}[theorem]{Corollary}

\makeatletter
\@addtoreset{equation}{section}

\makeatother

\subheader{{\sf\hfill YGHP-21-3}}
\title{The moduli space of non-Abelian vortices in
  Yang-Mills-Chern-Simons-Higgs theory} 

\author{Sven Bjarke Gudnason$^1$,}
\affiliation{$^1$Institute of Contemporary Mathematics, School of
  Mathematics and Statistics, Henan University, Kaifeng, Henan 475004,
  P.~R.~China}
\emailAdd{gudnason(at)henu.edu.cn}
\author{Minoru Eto$^{2,3}$}
\affiliation{$^2$Department of Physics, Yamagata University,
  Kojirakawa-machi 1-4-12, Yamagata, Yamagata 990-8560, Japan}
\affiliation{$^3$Department of Physics, Keio University, 4-1-1
  Hiyoshi, Kanagawa 223-8521, Japan}
\emailAdd{meto(at)sci.kj.yamagata-u.ac.jp}

\abstract{
We determine the dimension of the moduli space of non-Abelian vortices
in Yang-Mills-Chern-Simons-Higgs theory in $2+1$ dimensions for gauge
groups $G=(\U(1)\times G')/\mathbb{Z}_{n_0}$ with $G'$ being
an arbitrary semi-simple group and $n_0$ the greatest common
divisor of the Abelian charges of the $G'$ invariants.
The calculation is carried out using a Callias-type index
theorem, the moduli matrix approach and a D-brane setup in Type IIB
string theory.
We prove that the index theorem gives the number of zeromodes or
moduli of the non-Abelian vortices, extend the moduli matrix approach
to the Yang-Mills-Chern-Simons-Higgs theory and finally derive
the effective Lagrangian of Collie and Tong using string theory.
}

\begin{document}
\maketitle

\section{Introduction}

Chern-Simons (CS) field theory is an interesting topological field theory
that is nondynamical. Once coupled with Yang-Mills (Maxwell) theory,
it becomes a nontrivial theory giving mass to the gauge field, without
resort to the Higgs mechanism \cite{Deser:1981wh}.
Once matter fields and the Higgs mechanism is included as well, the
gauge fields possess a complicated pole structure, with two mass poles
in their propagator.
Abelian Chern-Simons theory has proved its importance in the subject
of the fractional quantum Hall effect \cite{Zhang:1988wy}, whereas the
non-Abelian Chern-Simons term is crucial in the Witten-Sakai-Sugimoto model
\cite{Witten:1998zw,Sakai:2004cn} as it reproduces the chiral anomaly
of quantum chromodynamics (QCD) and the Wess-Zumino-Witten term of the
Skyrme model at low energies. 

In this paper, we shall focus on 2-dimensional solitons in gauge
theory, namely vortices. The prime example of such solitons are
Abrikosov vortices in Type-II superconductors and their relativistic
generalization called Nielsen-Olesen or Abelian-Higgs vortices, for
reviews see e.g.~refs.~\cite{Jaffe:1980,Manton:2004}.
The Chern-Simons vortices were first studied in the Abelian
Chern-Simons-Higgs theory \cite{Paul:1986ix} and
Yang-Mills-Chern-Simons-Higgs (YMCSH) theory, but with the scalar fields in 
the adjoint representation of $\SU(N)$, hence giving rise to
$\mathbb{Z}_N$ vortices \cite{deVega:1986eu,deVega:1986hm,Kumar:1986yz}.
An interesting discovery was that Abrikosov-Nielsen-Olesen vortices,
that are magnetic vortices and electrically neutral, become
electrically charged when the Chern-Simons term is introduced in the
theory.
This is a mere consequence of Gauss' law.
Abelian selfdual Chern-Simons-Higgs vortices were found later on with
a special sixth-order potential and a vortex
Bogomol'nyi-Prasad-Sommerfield (BPS) equation that has a fourth-order
scalar-field dependence \cite{Hong:1990yh,Jackiw:1990pr}.
Abelian selfdual Maxwell-Chern-Simons-Higgs vortices were then
discovered \cite{Lee:1990eq,Lee:1991td,Lee:1992yc}, where the new
ingredient is not a sixth-order potential as in the pure Chern-Simons
case, but a normal fourth-order potential and the addition of an extra 
neutral scalar field as well as an interaction potential between the
Higgs and the neutral scalar field.
Although non-Abelian theories with YMCSH
vortices were studied already in
refs.~\cite{deVega:1986eu,deVega:1986hm,Kumar:1986yz}, they are not
non-Abelian vortices with orientational moduli. The latter appear in
$\U(N)$ gauge theories (as opposed to $\SU(N)$ gauge theories), where
the crucial extra $\U(1)$ factor allows for the fundamental group
$\pi_1(G)\simeq\mathbb{Z}$, namely any number of vortices, whereas
$\SU(N)$ vortices has a fundamental group
$\pi_1(\SU(N)/\mathbb{Z}_N)\simeq\mathbb{Z}_N$.
The genuine type of non-Abelian vortex with orientational moduli was
first found in refs.~\cite{Hanany:2003hp,Auzzi:2003fs,Shifman:2004dr}
in the context of Yang-Mills-Higgs theory.
Such genuine non-Abelian Chern-Simons-Higgs vortices were found later
on in refs.~\cite{Aldrovandi:2007nb,Lozano:2007yz,NavarroLerida:2008uj} and
with both Yang-Mills and Chern-Simons terms
in refs.~\cite{Collie:2008mx,Collie:2008za}.
An important distinction between Yang-Mills-Higgs vortices and
Chern-Simons-Higgs vortices is that the former are topological
vortices, whereas the latter can be topological or nontopological.
This feature is in general expected to persist in
YMCSH theory, because of the existence of an
unbroken phase, although we are not aware of any explicit solutions of
nontopological vortices in this theory.
Nevertheless, in this paper we shall focus on the topological vortices
in YMCSH theory only and this means that we study
the theory in the gapped, asymmetric or broken phase.
For nice reviews on Chern-Simons vortices, see
e.g.~refs.~\cite{Dunne:1995,Yang:2004,Tarantello:2008,Horvathy:2008hd}. 

The genuine non-Abelian vortices discussed so far are all in $\U(N)$
gauge theories, which in some sense is the most natural choice for
this type of non-Abelian vortices with orientational
($\mathbb{C}P^{N-1}$) moduli.
The generalization of Yang-Mills-Higgs vortices from $\U(N)$ to
$(\U(1)\times G')/\mathbb{Z}_{n_0}$\footnote{Here $n_0$ is
    the greatest common divisor of the Abelian charges of the $G'$
    invariants.} theories was made in 
refs.~\cite{Ferretti:2007rp,Eto:2008yi,Eto:2009bg,Gudnason:2010rm,Eto:2010aj,Eto:2011cv} because
of the group theoretical nontriviality of the corresponding
non-Abelian monopoles that the vortex strings can end on, for example
in a cascading symmetry breaking theory considered in
refs.~\cite{Ferretti:2007rp,Gudnason:2010jq}. 
A special feature of non-Abelian vortices in theories with $G'$ being
a smaller group than $\SU(N)$, like for instance $\SO(N)$ or
$\USp(2M)$ (with $N=2M$), is that they generically contain semilocal
or size moduli \cite{Eto:2009bg} and that the moduli space contains
fractional vortices \cite{Eto:2009bg,Eto:2009bz}.
The parallel development in non-Abelian Chern-Simons-Higgs theory of
generalizing the gauge group from $\U(N)$ to $(\U(1)\times G')/\mathbb{Z}_{n_0}$ was made
in refs.~\cite{Gudnason:2009ut,Gudnason:2010yy,Eto:2010mu}. 

In this paper, we work out the general equations for the BPS vortices
in YMCSH theory and mainly focus on
determining the dimension of the moduli space, that is, the number of
zeromodes contained in solutions to these BPS equations.
In doing so, we address the problem at hand using a Callias-type index
theorem, the moduli matrix approach as well as a brane construction in
Type IIB string theory.
The index theorem calculation is a non-Abelianization
of the calculation carried out in ref.~\cite{Lee:1991td}, however,
instead of splitting the equations up into real and imaginary parts,
we retain the complex structure, which is crucial for the next step.
The index only calculates the difference between the number of
zeromodes of the operator that corresponds to the fluctuation
equations of the BPS equations and the number of zeromodes of its
adjoint.
This only gives a lower bound on the number of zeromodes, unless we
can prove that the adjoint operator does not possess normalizable
zeromodes (which was the next step referred to above).

A main result of this paper, is the proof that the adjoint
operator does indeed not possess any normalizable zeromodes, which
is established by theorems \ref{thm:1} and \ref{thm:2} and their
corresponding proofs.
The proof partially uses the vacua-in-gauge-theory-approach
utilized in refs.~\cite{Hanany:2003hp,Eto:2009bg}, but a crucial point
was to find a useful gauge fixing condition for the fluctuation
spectrum.
The rest of the proof uses a Hermitian constraint on the fluctuations
of the gauge fields, which on top of the BPS vortex background
solutions is strong enough to eliminate the possibilities of
fluctuations of the adjoint operator.
After establishing the number of zeromodes, which is also the
dimension of the moduli space (of vortex solutions), we rewrite the
BPS equations into a master form using the moduli matrix approach.
The moduli matrix approach is essentially the non-Abelian solution to
the selfdual equation that is shared among most BPS solitons and fully
determines the non-Abelian gauge field in terms of a moduli matrix and
a residual field that does not contain moduli.
This approach was used for non-Abelian domain walls
\cite{Isozumi:2004jc}, composite solitons \cite{Isozumi:2004vg}, and
non-Abelian vortices \cite{Eto:2005yh} in the $\U(N)$ Yang-Mills-Higgs
theory, see ref.~\cite{Eto:2006pg} for a review.
It was also used in
refs.~\cite{Eto:2008yi,Eto:2009bg,Eto:2009bz} where it was
applied to the non-Abelian vortices in Yang-Mills-Higgs theory and in
refs.~\cite{Gudnason:2009ut,Gudnason:2010yy,Eto:2010mu} in 
non-Abelian Chern-Simons-Higgs theory, both with gauge groups
$(\U(1)\times G')/\mathbb{Z}_{n_0}$.
All these well-studied vortices are included in our BPS theory as
several limits. For example, the non-Abelian vortices in
Yang-Mills-Higgs theory \cite{Eto:2005yh} are obtained in the weak
gauge coupling limit (and setting the adjoint scalar field to zero). 
The non-Abelian Chern-Simons-Higgs vortex \cite{Gudnason:2009ut}, on the
other hand, is obtained by taking the strong gauge coupling limit.
In this sense, the theory at hand is the most generic theory and we
derive in this paper the most generic form of the moduli matrix method
for the non-Abelian vortices.

The final approach to the moduli space of non-Abelian vortices in
YMCSH theory is a generalization of the brane construction by Hanany
and Tong \cite{Hanany:2003hp} from having only Yang-Mills theory to
having both Yang-Mills and Chern-Simons terms in the theory on the
brane (the vortex). 
The D-brane construction of supersymmetric 3-dimensional Chern-Simons
theory in Type IIB string theory has been studied in
refs.~\cite{Kitao:1998mf,Ohta:1999gj,Bergman:1999na}, which involves
studying so-called $(p,q)$-branes, where the Chern-Simons level $\kappa$
is related to the ratio $p/q$.
Abelian Chern-Simons vortices were studied in this context in
refs.~\cite{Ohta:1999gj,Lee:1999ze}.
T-duality can be used to make the brane construction of non-Abelian
vortices in Type IIA string theory \cite{Hanany:2004ea} instead of in
Type IIB string theory \cite{Hanany:2003hp} and the same holds true
also for the Abelian vortices in Chern-Simons theory
\cite{Brodie:2000ns}. 
Furthermore, the D-brane construction also allows us to write
down the effective theory on the world-line, exactly giving rise to the
result obtained in ref.~\cite{Collie:2008mx}.

The organization of the paper is as follows.
In sec.~\ref{sec:model}, we introduce YMCSH
theory, its various limiting theories, it mass spectrum and the BPS
equations for vortices.
In sec.~\ref{sec:indextheorem}, we calculate the index of the operator
corresponding to the zeromodes of the vortices and prove that this
index exactly counts this number.
In sec.~\ref{sec:modulimatrix}, we extend the moduli matrix approach
to the YMCSH theory by including an extra adjoint field in the now
coupled master equations.
We also give some examples of vortex equations at the center of
orientational patches on standard Taubes-like equation form.
In sec.~\ref{sec:dbranepicture}, we construct the D-brane setup for
the vortices in Type IIB string theory, read off the number of
zeromodes and its low-energy effective Lagrangian from the vortex
brane.

\section{The model}\label{sec:model}

We consider ${\cal N}=2$ supersymmetric
Maxwell-Yang-Mills-Chern-Simons-Higgs (YMCSH) theory in $d=2+1$ dimensions  
with the gauge group $G=(\U(1)\times G')/\mathbb{Z}_{n_0}$ and often
take $G'$ to be $\SU(N)$, but will keep it general in most equations.
$n_0$ is the greatest common divisor of the Abelian charges of the
invariants of $G'$.
In particular, for $G'=\SU(N)$ it is $n_0=N$, for
$G'=\SO(2M),\USp(2M)$ it is $n_0=2$ and for $G'=\SO(2M+1)$ it is
$n_0=1$.
$M\in\mathbb{Z}_{>0}$ is taken to be a non-negative integer, whereas
$N$ is always the rank of $G$. 

The bosonic part of the Lagrangian density reads
\begin{align}
\mathcal{L} = &
-\frac{1}{4g^2}(F_{\mu\nu}^a)^2
-\frac{1}{4e^2}(F_{\mu\nu}^0)^2
-\frac{\mu}{8\pi}\epsilon^{\mu\nu\rho}
  \left(A_{\mu}^a\partial_\nu A_\rho^a  
  -\frac{1}{3}f^{abc}A_{\mu}^a A_{\nu}^b A_{\rho}^c\right)
-\frac{\kappa}{8\pi}
  \epsilon^{\mu\nu\rho} A_{\mu}^0\partial_\nu A_\rho^0 \non
&+\frac{1}{2g^2}(D_\mu\phi^a)^2
+\frac{1}{2e^2}(\partial_\mu\phi^0)^2
+\tr\big[(D^\mu H)(D_\mu H)^\dag\big]
-\tr\big[(\phi H - H m)(\phi H - H m)^\dag\big]\non
&-\frac{g^2}{2}\left(\tr[HH^\dag t^a]-\frac{\mu}{4\pi}\phi^a\right)^2
-\frac{e^2}{2}\left(\tr[HH^\dag t^0]-\frac{\kappa}{4\pi}\phi^0
  -\frac{\xi}{\sqrt{2N}}\right)^2, \label{eq:LYMCSH}
\end{align}
where $a=1,2,\ldots,\dim(G')$ is the gauge index for $G'$ and the index
$0$ denotes the $\U(1)$ gauge group. The gauge couplings $e,g$ are for 
the $\U(1)$ and the $G'$ part of the Yang-Mills term, respectively. 
Similarly, the two Chern-Simons couplings $\kappa,\mu$ are for the
$\U(1)$ and the $G'$ part of the Chern-Simons term,
respectively, and take values in real and integer numbers:
  $\kappa\in\mathbb{R}$ and $\mu\in\mathbb{Z}$. The integral condition
  on the coupling renders the non-Abelian Chern-Simons term invariant
  under large gauge transformations \cite{Deser:1981wh}. 
The $\NF=N$ complex scalar fields $H$ are in the fundamental
representation of $G'$ and are therefore $N\times\NF$ complex matrices
and the real scalar field $\phi^a$ is in the adjoint representation of
$G'$ ($\phi^0$ is a singlet).
The mass matrix $m$ is a complex $\NF\times\NF$ matrix which gives
rise to the mass-deformed version of the theory at hand (see
e.g.~ref.~\cite{Eto:2011cv} for a mass-deformed version of
Yang-Mills-Higgs theory with arbitrary gauge groups), but in this
paper we will consider only the case of $m=0$.
We fix our conventions as
\begin{align}
F_{\mu\nu} &= \p_\mu A_\nu - \p_\nu A_\mu + \i[A_\mu,A_\nu],\\
D_\mu H &= (\p_\mu + \i A_\mu)H,\\
D_\mu\phi &= \p_\mu\phi + \i[A_\mu,\phi],
\end{align}
with $\phi = \phi^\alpha t^\alpha$ and $A_\mu = A_\mu^\alpha t^\alpha$. 
Here we choose a standard notation
\beq
\tr(t^\alpha t^\beta) = \frac{1}{2}\delta^{\alpha\beta}, \quad 
t^0 = \frac{{\bf 1}_N}{\sqrt{2N}},
\label{eq:tt_normalization}
\eeq
for $\alpha,\beta = 0,1,2,\ldots,\dim(G')$, which are gauge group
indices running over both $\U(1)$ and $G'$.
The so-called Fayet-Illiopoulos (FI) parameter $\xi>0$ is positive,
which ensures stable supersymmetric vacua.
To simplify the notation, we define the following parameter
\beq v \equiv \sqrt{\frac{\xi}{N}} > 0. \eeq

Let us briefly discuss the spectrum of the theory in various limits in
which the above theory reduces to other well-known theories. 
\begin{itemize}
\item Maxwell-Yang-Mills-Higgs ($\kappa\to0$ and $\mu\to 0$, $m=0$): 

There exists a unique Higgs vacuum $H = v{\bf 1}_N$ and $\phi=0$. The
mass spectrum is
\beq
m_e = e v, \quad
m_g = g v,
\label{eq:higgs_mass}
\eeq
which are the masses of the Abelian and non-Abelian perturbative
excitations of the fields, respectively. 
Due to supersymmetry, the vector multiplets and the chiral multiplets
have the same masses. Namely, the Abelian gauge field $A_\mu^0$, the
real scalar field $\phi^0$ and the real part of the 
trace part of $H$ all have the same mass $m_e$. 
On the other hand, the $G'$ gauge fields $A_\mu^a$, the real
adjoint fields $\phi^a$ and the real part of the traceless part of
$H\in\{H|\tr(H H^\dag t^a)\neq0\}$ all have the mass $m_g$.  
The vortex in this model is a non-Abelian $G$-vortex which is a
natural extension of the Nielsen-Olesen vortex of the Abelian-Higgs 
model. 
The $G$-vortex has internal orientational modes associated with the
spontaneously broken global symmetry in addition to the zeromode
corresponding to the center of mass position.
In the $G'=\SU(N)$ case, these zeromodes of a single non-Abelian
vortex live on $\mathbb{C}P^{N-1}$, whereas for $G'=\SO(2M)$
and $G'=\USp(2M)$, ($N=2M$) they live on $\SO(2M)/\U(M)$ and
$\USp(2M)/\U(M)$, respectively.

\item Maxwell-Yang-Mills-Chern-Simons ($H:=0$, $m=0$):

If we switch off the Higgs fields $H$ from the full Lagrangian
\eqref{eq:LYMCSH}, the model reduces to the
Maxwell-Yang-Mills-Chern-Simons model. 
The FI term in eq.~\eqref{eq:LYMCSH} can be absorbed by a shift of
$\phi^0$, which is equivalent to setting $\xi=0$.  
The vacuum is in the symmetric phase with $\phi=0$. No symmetries
are broken. Although the Higgs mechanism is not at work, the vector
multiplet acquires a topological mass due to the Chern-Simons term. 
The masses are given by
\beq
m_\kappa = \frac{\kappa e^2}{4\pi}, \quad
m_\mu = \frac{\mu g^2}{4\pi}.
\label{eq:topological_mass}
\eeq
$m_\kappa$ is the mass of the Abelian gauge field $A_\mu^0$ and also
the real scalar field $\phi^0$. 
$m_\mu$ is the mass of $G'$ gauge field $A_\mu^a$ as well as the real 
adjoint field $\phi^a$.

\item Chern-Simons-Higgs ($e\to\infty$ and $g\to\infty$, $m=0$):

The Maxwell and Yang-Mills kinetic terms \eqref{eq:LYMCSH} disappear
in this limit.
Hence, the Lagrangian \eqref{eq:LYMCSH} reduces to the
Chern-Simons-Higgs model (see e.g.~ref.~\cite{Gudnason:2009ut}). 
Furthermore, the adjoint scalar fields $\phi^\alpha$ become
nondynamical fields and can be integrated out as follows
\beq \phi^0 = \frac{4\pi}{\kappa}\bigg(
\tr(HH^\dag t^0) - \sqrt{\frac{N}{2}}v^2 \bigg), \quad
\phi^a = \frac{4\pi}{\mu}\tr(HH^\dag t^a). \eeq
Plugging these back into eq.~\eqref{eq:LYMCSH}, one obtains the
sixth-order scalar potential 
\beq
V_{e,g\to\infty} = \tr[\phi^2 HH^\dag].
\eeq
There exist three types of vacua in this model. One is in the
completely symmetric phase where $H=0$ and no symmetries are broken. 
The vector multiplets are decoupled, so the Higgs fields are the only 
dynamical degrees of freedom. 
Their masses are
\beq
m_H = |\phi^0| = \frac{4\pi v^2}{\kappa} \sqrt{\frac{N}{2}}.
\eeq
In between there is a variety of partially broken phases, which we
will not discuss in detail here.
Finally, there is the vacuum in the Higgs phase, where the gauge
symmetry is completely broken. 
The longitudinal part of the gauge field acquires a mass through the
Higgs mechanism. 
Moreover, the mass of the Higgs field is the same as that of the gauge
field due to supersymmetry
\beq
m_{\kappa\infty} = \frac{m_e^2}{m_\kappa},\quad
m_{\mu\infty} = \frac{m_g^2}{m_\mu}.
\label{eq:topological_mass2}
\eeq
In the symmetric phase of the $N=1$ (Abelian) case, it is known that
nontopological vortices exist. 
On the other hand, there exist topological solitons in the broken
phase in the Abelian \cite{Hong:1990yh,Jackiw:1990pr} as well as the
non-Abelian theory \cite{Aldrovandi:2007nb,Lozano:2007yz} (see also
refs.~\cite{deVega:1986hm,Kumar:1986yz,deVega:1986eu,NavarroLerida:2008uj}
for a non-Abelian variant of the theory with the matter fields in the
adjoint representation).
\end{itemize}

The topological and non-topological vortices in the
Maxwell-Chern-Simons-Higgs model have also been paid great attention
in the literature \cite{Lee:1990eq,Lee:1991td,Lee:1992yc}
($N=1$, $\mu=g=0$ in Lagrangian \eqref{eq:LYMCSH}). 
Its non-Abelian generalization has also been studied
\cite{Collie:2008mx,Collie:2008za}:
\begin{itemize}
\item $\kappa=\mu$, $e=g$ and $m=0$: The
  Maxwell-Yang-Mills-Chern-Simons vortex. 
\item $\kappa=\mu$, $e=g$ and $m\neq0$: The dyonic non-Abelian
  Chern-Simons vortex.
\end{itemize}


In the following we will set 
\beq
m=0,
\eeq
and set the number of flavors $\NF:=N$ so that $H$ is a complex square 
matrix. 

Now we are ready to identify the mass spectrum in the full Lagrangian
\eqref{eq:LYMCSH} with the coupling constants kept at generic finite
(and nonvanishing) values. There are two ground states respecting the
full supersymmetry of the theory.  
The first one is an unbroken phase (i.e.~the symmetric phase) and the
second one is the Higgs phase (i.e.~the broken phase). The unbroken
phase is given by 
\beq
\text{unbroken phase}:\qquad
\phi^0 = - \frac{4\pi v^2}{\kappa} \sqrt{\frac{N}{2}}, \quad
\phi^a = 0,\quad
H = 0.
\label{eq:vac_ub}
\eeq
The gauge and flavor symmetries are not broken in this phase. Although
the gauge symmetry is not broken, the gauge fields acquire
topologically induced masses.
The Abelian gauge fields and the real scalar field $\phi^0$ have the
same mass $m_\kappa$, while the $G'$ gauge fields and the adjoint scalar 
$\phi^a$ have the same mass $m_\mu$. This mass degeneracy is of course
due to ${\cal N}=2$ supersymmetry. 

On the other hand, the gauge symmetry is broken in the Higgs phase
(i.e.~the asymmetric or broken phase) where the fields develop the
following vacuum expectation values (VEVs):
\beq
\text{broken phase}:\qquad
\phi^0 = 0,\quad \phi^a = 0, \quad
H = v {\bf 1}_N.
\label{eq:vac_b}
\eeq
The symmetry breaking pattern of the continuous part of the
  symmetry is:
\beq
\U(1)_{\rm c}\times G' \times\SU(N)_{\rm f} \to G'_{\rm c+f},
\label{eq:cf}
\eeq
hence the broken phase is in the color-flavor locking phase and this
was the main reason for setting $\NF=N$.
Explicitly, for $G'=\SU(N)$ we have
\beq
\U(1)_{\rm c}\times\SU(N)_{\rm c}\times\SU(N)_{\rm f} \to \SU(N)_{\rm c+f}.
\nonumber
\eeq
Let us now identify the mass spectrum in the broken phase. 
Note that $1+\dim G'$ real degrees of freedom in $H$ are eaten by the gauge
fields via the Higgs mechanism, which is $N^2$ real degrees of freedom 
for $G'=\SU(N)$ and exactly half of the real degrees of freedom of
$H$.
For other classical continuous groups, like $G'=\SO(N)$ and
$G'=\USp(2M)$, it is smaller: $N(N-1)/2$ and $M(1+2M)$, respectively. 
The gauge fields aquire their masses through both the Chern-Simons
topological mechanism and the Higgs mechanism. Their mass poles of
propagators of the Abelian and the $G'$ gauge fields split into two
poles for each propagator respectively as (see
ref.~\cite{Pisarski:1985yj,Paul:1985mn} for the Abelian case)
\beq
m_{e\kappa\pm} = \sqrt{m_e^2+\frac{m_\kappa^2}{4}} 
  \pm \frac{m_\kappa}{2}, \quad
m_{g\mu\pm} = \sqrt{m_g^2+\frac{m_\mu^2}{4}} \pm \frac{m_\mu}{2},
\label{eq:gene_mass}
\eeq
where $m_e,m_g$ are given in eq.~\eqref{eq:higgs_mass} and
$m_\kappa,m_\mu$ are given in eq.~\eqref{eq:topological_mass}.
Note that we can reproduce the topological masses of
eq.~\eqref{eq:topological_mass2} as the limit where the couplings $e$ 
and $g$ are sent to infinity 
\beq
\lim_{e\to\infty} m_{e\kappa-} = m_{\kappa\infty}, \quad
\lim_{e\to\infty} m_{e\kappa+} = \infty, \quad
\lim_{g\to\infty} m_{e\mu-} = m_{\mu\infty}, \quad
\lim_{g\to\infty} m_{e\mu+} = \infty,
\eeq
where in each mass, one pole is sent off to infinity and the other
pole converges to the topological mass. 

In order to identify the masses of the scalar fields, let us consider
the following small fluctuations\footnote{Note that if $G'$
    is a subgroup of $\SU(N)$, then all the complimentary elements
    to $G'$ of the matrix $H$ remain massless, whereas for $G'=\SU(N)$
    all elements of $H$ acquire a mass.}
\beq
H = v{\bf 1}_N + 
\left(\delta H_1^\alpha + \i\delta H_2^\alpha \right)t^\alpha, \quad
\phi = -e \delta \phi^0 t^0 - g \delta \phi^a t^a.
\eeq
Let us furthermore define 
$\delta\Phi^\alpha\equiv\left(\delta H_1^\alpha,\ \delta \phi^\alpha
\right)$. 
Then the quadratic terms in the fluctuations of the Lagrangian in the
scalar sector are found to be 
\begin{equation}
{\cal L}^{(2)} = \frac{1}{2}\p_\mu \delta H_2^\alpha \p^\mu \delta H_2^\alpha
+ \frac{1}{2} \p_\mu \delta \Phi^\alpha \p^\mu (\delta \Phi^{\alpha})^{\rm T}
- \frac{1}{2} \delta \Phi^0 M_e^2 (\delta \Phi^0)^{\rm T}
- \frac{1}{2} \delta \Phi^\alpha M_g^2 (\delta \Phi^\alpha)^{\rm T},
\end{equation}
where we have introduced the following Hermitian mass-squared matrices 
\beq
M_e^2 \equiv 
\begin{pmatrix}
m_e^2 & m_e m_\kappa\\
m_e m_\kappa & m_e^2 + m_\kappa^2
\end{pmatrix}, \quad
M_g^2 \equiv 
\begin{pmatrix}
m_g^2 & m_g m_\mu\\
m_g m_\mu & m_g^2 + m_\mu^2
\end{pmatrix}.
\label{eq:mass_matrix}
\eeq
It is then possible to read off the mass spectrum by diagonalization
of these mass-squared matrices. For later convenience, let us rewrite
the mass matrices in the following way 
\beq
M_e^2 = U_e 
\begin{pmatrix}
m_{e\kappa-}^2 & \\
& m_{e\kappa+}^2
\end{pmatrix}
U_e^{-1}, \quad
M_g^2 = U_g 
\begin{pmatrix}
m_{g\mu-}^2 & \\
& m_{g\mu+}^2
\end{pmatrix}
U_g^{-1},
\eeq
where the $\SO(2)$ matrices $U_{e,g}$ are defined as follows
\beq
U_{e,g} = 
\begin{pmatrix}
\cos u_{e,g} & \sin u_{e,g}\\
-\sin u_{e,g} & \cos u_{e,g}
\end{pmatrix}, \quad
\tan 2u_{e,g} = \frac{2 m_{e,g}}{m_{\kappa,\mu}}, \quad
u_{e,g}\in\big[0,\tfrac{\pi}{4}\big].
\eeq

Again the degeneracy between the masses of the vector multiplets and 
the chiral multiplets occurs because the Lagrangian respects ${\cal N}=2$
supersymmetry. Hence we have four different characteristic mass scales
in our system. 
Let us introduce the following definition
\beq
\gamma_{i/j} = \frac{m_i^2}{m_j^2},\quad
i,j=e,g,\mu,\kappa,e\kappa\pm,g\mu\pm,
\label{eq:gammas}
\eeq
which are dimensionless parameters characterizing the solutions.
However, only three of these parameters $\gamma_{i/j}$ are
independent. 


We are interested in vortex-type solitons in the Higgs vacuum.
Since $(\U(1)_{\rm c}\times G')/\mathbb{Z}_{n_0}$ is
spontaneously broken, we expect 
topologically stable vortices, which are supported by  
\beq
\mathbb{Z} = \pi_1\big(\U(1)_{\rm c}\big)
\subset \pi_1\big((\U(1)_{\rm c}\times G')/\mathbb{Z}_{n_0}\big).
\eeq
Thanks to ${\cal N}=2$ supersymmetry, the vortices in this system are
of the Bogomol'nyi-Prasad-Sommerfield (BPS) type of soliton, which
preserves half of supersymmetry. Static interactions between multiple
vortices are completely canceled out, hence we can construct static
multi-vortices.
Instead of solving half of the supersymmetry-preserving conditions on
the fermions, we can derive the BPS equations by performing a standard
Bogomol'nyi completion of the energy functional 
\begin{align}
T = \int_{\mathbb{C}} \Bigg(&
\frac{1}{2g^2}\left[F_{12}^a - 
  g^2\left(\tr(H H^\dag t^a) - 
  \frac{\mu}{4\pi}\phi^a\right)\right]^2
+\frac{1}{2g^2}[F_{0i}^a + D_i\phi^a]^2 \\&
+\frac{1}{2e^2}\left[F_{12}^0 - 
  e^2\left(\tr(H H^\dag t^0) - 
  \frac{\kappa}{4\pi}\phi^0 - \frac{\xi}{\sqrt{2N}}\right)\right]^2 
+\frac{1}{2e^2}[F_{0i}^0 + \p_i\phi^0]^2 \non&
+4\tr|D_{\bar{z}} H|^2
+\tr|D_0 H - \i\phi H|^2
+\frac{1}{2g^2}(D_0\phi^a)^2
+\frac{1}{2e^2}(\p_0\phi^0)^2 \non&
-\frac{\xi}{\sqrt{2N}}F_{12}^0
-\frac{1}{g^2}\p_i(F_{0i}^a\phi^a)
-\frac{1}{e^2}\p_i(F_{0i}^0\phi^0)
-\i\epsilon^{ij}\tr\p_i\big((D_j H)H^\dag\big)
\Bigg)\;\d^2x, \nonumber
\end{align}
where $i,j=1,2$, we have used Gauss' law \eqref{eq:Gausslaw} and
we have defined
\beq
D_{\bar{z}} = \frac{D_1 + \i D_2}{2}.
\eeq
The tension is bounded from below by the Bogomol'nyi bound
\begin{align} 
  T \ge T_{\rm BPS}
  = -\frac{\xi}{\sqrt{2N}}\int_{\mathbb{C}}F_{12}^0\;\d^2x
  = 2\pi\xi\nu = \frac{2\pi\xi k}{n_0},\quad
  k \in \mathbb{Z}_{\ge 0},
\end{align}
with 
\begin{align}
\nu = - \frac{1}{2\pi\sqrt{2N}}\int_{\mathbb C} F_{12}^0\;\d^2x = \frac{k}{n_0}, 
\end{align}
being the $U(1)$ winding number and $n_0$ is the greatest common
divisor of the Abelian charges of the $G'$ invariants.
This result was establish in ref.~\cite{Eto:2008yi} and the analysis
looks at the winding of all $G'$ invariants at spatial infinity.
The smallest ``closed loop'' is made by having a $\U(1)$ winding of
$1/n_0$ and returning inside the group $G'$.
Any smaller winding numbers would thus make some invariants
non-single-valued at infinity.
Due to the normalization of the $\U(1)$ gauge fields
\eqref{eq:tt_normalization}, the factor of $1/\sqrt{2N}$ appears
explicitly in the definition of the $\U(1)$ winding number.
For $G'=\SU(N)$, $n_0=N$ and the BPS tension is simply
$T_{\rm BPS}=2\pi\xi k/N=2\pi v^2 k$. 

The inequality is saturated when the BPS equations are satisfied:
\begin{align}
&D_{\bar{z}} H = 0, \\
  &\hat F_{12} - g^2\left(\langle H H^\dag\rangle_{G'}
  - \frac{\mu}{4\pi}\hat\phi\right) = 0, \\
&F_{12}^0 - 
  e^2\left(\frac{1}{\sqrt{2N}}\tr(H H^\dag) - 
  \frac{\kappa}{4\pi}\phi^0 - \frac{\xi}{\sqrt{2N}}\right) = 0, \\
&F_{0i} + D_i\phi = 0,\quad
\left(D_0  - \i\phi\right) H = 0, \quad
D_0\phi = 0,
\end{align}
where we have introduced the notation
\beq
\hat{\phi} = \phi^a t^a,
\eeq
for the $G'$ part of fields (and similarly for $F_{12}$), as well as
the projection operator
\beq
\langle X\rangle_{G'} = \tr[X t^a] t^a,
\eeq
for an arbitrary matrix $X$. 
Note that the above BPS equations are to be accompanied by Gauss' law
\begin{align}
\frac{1}{g^2}D_i \hat F_{0i}
-\frac{\mu}{4\pi} \hat F_{12}
+\frac{\i}{g^2}\big[\hat \phi,D_0 \hat \phi\big] = \hat j_0,\quad
\frac{1}{e^2}\p_i F_{0i}^0
-\frac{\kappa}{4\pi}F_{12}^0
= j_0^0,
\label{eq:Gausslaw}
\end{align}
where we have defined the Noether current
\beq
j_0 = j_0^\alpha t^\alpha \equiv
\i\left\langle(D_0 H) H^\dag - H(D_0 H)^\dag\right\rangle_{G}, 
\eeq
and the $G$ projection operator
\beq
\langle X\rangle_G \equiv \tr(X t^\alpha)t^\alpha
= \langle X\rangle_{G'} + \frac{1}{\sqrt{2N}}\tr(X)\mathbf{1}_N,
\eeq
and the covariant derivative acts on the field strength as
\beq
D_i\hat{F}_{0i} = \p_i\hat{F}_{0i} + \i\big[A_i,\hat{F}_{0i}\big].
\eeq
Anti-BPS vortices can be obtained in similar way by a different choice
of signs in writing the squares when doing the Bogomol'nyi
completion. 
An important feature of the vortices in Chern-Simons theory is that
they carry electric charges. This fact can be seen directly from
Gauss' law. The electric charge is fractionally quantized since it is
related to the magnetic flux as
\beq 
Q \equiv \frac{1}{\sqrt{2N}}\int_{\mathbb{C}} j_0^0 \;\d^2x
  = -\frac{\kappa}{4\pi\sqrt{2N}} \int_{\mathbb{C}} F_{12}^0\;\d^2x
= \frac{\kappa k}{2n_0}.
\eeq

In order to simplify the BPS equations, we can consistently choose
$\phi = A_0$ in temporal gauge. Moreover, we will only consider static
solutions in this paper. Hence, the BPS equations combined with Gauss'
law yield the following system of equations
\begin{align}
&D_{\bar{z}} H = 0, \label{eq:solvable}\\
&\hat F_{12}
  = g^2\left(\langle H H^\dag\rangle_{G'}
  -\frac{\mu}{4\pi}\hat\phi\right), \quad
F_{12}^0 
 = \frac{e^2}{\sqrt{2N}}\bigg(\tr(H H^\dag)
  -\frac{\kappa\sqrt{2N}}{4\pi}\phi^0 - \xi\bigg),
  \label{eq:master1}\\
&\frac{1}{g^2}D_i^2 \hat \phi
  +\frac{\mu}{4\pi} \hat F_{12}
 =\big\langle\{H H^\dag,\phi\}\big\rangle_{G'}, \quad
\frac{1}{e^2}\p_i^2\phi^0
  +\frac{\kappa}{4\pi}F_{12}^0 
  = \sqrt{\frac{2}{N}} \tr(H H^\dag\phi).
  \label{eq:master2}
\end{align}
For later convenience we write eq.~\eqref{eq:master2} as follows
\begin{align}
4D_z D_{\bar{z}} \hat\phi &= m_\mu^2\hat\phi
- g^2m_\mu\langle H H^\dag\rangle_{G'}
- g^2\big[\langle H H^\dag\rangle_{G'},\hat\phi\big]
+ g^2\big\langle\{ H H^\dag,\phi\}\big\rangle_{G'},\non
\p_i^2\phi^0 &= m_\kappa^2\phi^0
+\frac{e^2}{\sqrt{2N}}\tr\big[(2\phi-m_\kappa\mathbf{1}_N)H H^\dag\big]
+\sqrt{\frac{N}{2}}m_e^2m_\kappa,
\label{eq:master3}
\end{align}
and the former simplifies for $G'=\SU(N)$ to
\beq
4D_z D_{\bar{z}} \hat\phi = m_\mu^2 \hat\phi 
+ g^2\big\langle(2\phi-m_\mu{\bf 1}_N)H H^\dagger\big\rangle_{\SU(N)},
\label{eq:master3_SUN}
\eeq
where we have used the identity
$D_i^2\hat\phi =4D_z D_{\bar{z}}\hat\phi+[\hat{F}_{12},\hat\phi]$, and 
eq.~\eqref{eq:master1}.

The equations simplify for the $G=\U(N)$ case, by which we mean that
$G'=\SU(N)$ and the Abelian couplings equal their non-Abelian
counterparts: $e=g$ and $\kappa=\mu$.
In that case we can write
\begin{align}
  D_{\bar{z}} H &= 0, \label{eq:master1_UN}\\
  F_{12} &= \frac{g^2}{2} H H^\dag - m_\mu \phi - \frac12m_g^2\mathbf{1}_N, \label{eq:master2_UN}\\
  4D_z D_{\bar{z}} \phi &= m_\mu^2\phi
  + g^2\left(\phi-\frac12m_\mu\mathbf{1}_N\right)H H^\dag
  +\frac12m_g^2m_\mu\mathbf{1}_N. \label{eq:master3_UN}
\end{align}


\section{The index theorem}\label{sec:indextheorem}

We will now turn to the fluctuation spectrum around an arbitrary BPS
solution in the YMCSH theory.
In order to avoid a cluttered notation, we will carry out the
calculation for the $G=\U(N)$ case, but the $G'$ dependence in the end
result will remain transparent, as we will see shortly. 

In order to get a real equation for $\phi$ and its corresponding
fluctuations, we will add the Hermitian conjugate of the equation to
itself and divide by four, obtaining:
\begin{equation}
D_{\{\bar{z}}D_{z\}}\phi
=D_{\bar{z}}D_z\phi + D_z D_{\bar{z}}\phi
=\frac{m_\mu^2}{2}\phi
+\frac{g^2}{4}\{\phi,H H^\dag\}
-\frac{m_\mu g^2}{4} H H^\dag
+\frac{m_\mu m_g^2}{4}\mathbf{1}_N, \label{eq:YMCS-BPS3}
\end{equation}
where the curly brackets on the indices mean symmetrization.

It will prove convenient and simplify the notation, if we do the
following rescaling
\beq
z\to \frac{z}{m_g}, \qquad
\p_z\to m_g\p_z, \qquad
A_z\to m_g A_z, \qquad
H\to v H, \qquad
\phi\to\frac{m_g^2}{m_\mu}\phi,
\label{eq:rescaling}
\eeq
for which the BPS system of equations reads
\begin{align}
  D_{\bar{z}}H &= 0,\\
  F_{12} &= \frac12 H H^\dag - \phi - \frac12\mathbf{1}_N,\label{eq:BPS2_rescaled}\\
  D_{\{\bar{z}}D_{z\}}\phi &= \frac{\tau}{2}\phi
  +\frac14\{\phi,H H^\dag\} -\frac{\tau}{4} H H^\dag + \frac{\tau}{4}\mathbf{1}_N,
\end{align}
where we have defined
\beq
\tau\equiv \frac{m_\mu^2}{m_g^2}.
\label{eq:tau}
\eeq
Varying the left-hand-side of eq.~\eqref{eq:YMCS-BPS3} yields
\begin{align}
  \delta[D_{\{\bar{z}}D_{z\}}\phi] &=
  D_{\{\bar{z}}D_{z\}}\delta\phi
  +\i[D_{\{\bar{z}}\dA_{z\}},\phi]
  +\i 2[\dA_{z},D_{\bar{z}}\phi]
  +\i 2[\dA_{\bar{z}},D_z\phi],
\end{align}
where $\dphi$, $\dA_z$ and $\dA_{\bar{z}}$ denote the perturbations
(fluctuations) of the fields $\phi$, $A_z$ and $A_{\bar{z}}$,
respectively, and the fluctuations still have to obey the BPS
equations on the background of a given solution where they describe
moduli (zeromodes) of the solution.
The covariant derivative acts on the gauge field (which is adjoint
valued), and hence on the fluctuations of the gauge field, as
\beq
D_{\bar{z}}\dA_z = \p_{\bar{z}}\dA_z + \i\big[A_{\bar{z}},\dA_z\big].
\eeq
To perform the calculation, we need to fix the gauge of the
fluctuations and we choose
\beq
D_{\{\bar{z}}\dA_{z\}} + \dH X + Y \dH^\dag = 0,
\label{eq:gauge_fixing_cond}
\eeq
where $X$ and $Y$ are unspecified functions of the background fields
(i.e.~solutions to the BPS equations).

The linear fluctuation equations thus read
\begin{align}
\i D_{\bar{z}}\dH -\dA_{\bar{z}}H &= 0, 
\label{eq:YMCS-linearfluct1}\\
\i D_{\bar{z}}\dA_{z} - \i D_z\dA_{\bar{z}} - \frac14\dH H^\dag 
-\frac14 H \dH^\dag + \frac12\delta\phi &= 0, \label{eq:YMCS-linearfluct2}\\
D_{\{\bar{z}}D_{z\}}\delta\phi
-\i[\dH X + Y\dH^\dag,\phi]
+\i 2\left[\dA_{z},D_{\bar{z}}\phi\right] 
+\i 2\left[\dA_{\bar{z}},D_z\phi\right] \non
\mathop+\frac{\tau}{4}(\dH H^\dag + H\dH^\dag)
-\frac{\tau}{2}\delta\phi 
-\frac14\{\dH H^\dag,\phi\}
-\frac14\{H \dH^\dag,\phi\}
-\frac14\{H H^\dag,\delta\phi\} &= 0, \label{eq:YMCS-linearfluct3}
\end{align}
where we have used the gauge condition for the fluctuations
\eqref{eq:gauge_fixing_cond}, and in turn the operator equation is
\beq \mathbb{D} \eta = 0, \eeq
with the operator defined by
\begin{align}
\mathbb{D} &= \i D + K, \label{eq:decomp}\\
D &\equiv
\begin{pmatrix}
\p_{\bar{z}} & 0 & 0 & 0 & 0 \\
0 & \p_z & 0 & 0 & 0 \\
0 & 0 & -\frac{1}{\sqrt{2}}\p_z & \frac{1}{\sqrt{2}}\p_{\bar{z}} & 0 \\
0 & 0 & \frac{1}{\sqrt{2}}\p_z & \frac{1}{\sqrt{2}}\p_{\bar{z}} & 0 \\
0 & 0 & 0 & 0 & -\i 2D_{\bar{z}}\p_z -\i 2D_z \p_{\bar{z}} + \i 2\p_{\bar{z}}\p_z
\end{pmatrix}, \label{eq:Dop}\\
K &\equiv
\begin{pmatrix}
-A_{\bar{z}} & 0 & -\circ H & 0 & 0 \\
0 & \circ A_z & 0 & H^\dag & 0 \\
-\frac{1}{4\sqrt{2}} \circ H^\dag & -\frac{1}{4\sqrt{2}}H & 
  \frac{1}{\sqrt{2}}\left[A_z,\circ\right] & 
  -\frac{1}{\sqrt{2}}[A_{\bar{z}},\circ] & \frac{1}{2\sqrt{2}} \\
\circ X & Y & -\frac{1}{\sqrt{2}}[A_z,\circ] & 
  -\frac{1}{\sqrt{2}}[A_{\bar{z}},\circ] & 0 \\ 
K_{51} & K_{52} & K_{53} & K_{54} & K_{55}
\end{pmatrix},\\
K_{51} &\equiv \frac{\tau}{4}\circ H^\dag -\frac14\{\circ H^\dag,\phi\}
-\i[\circ X,\phi],\\
K_{52} &\equiv \frac{\tau}{4}H - \frac14\{H \circ,\phi\}
-\i[Y\circ,\phi],\\
K_{53} &\equiv \i 2[\circ,D_z\phi],\label{eq:K53}\\
K_{54} &\equiv \i 2[\circ,D_{\bar{z}}\phi],\label{eq:K54}\\
K_{55} &\equiv -\frac{\tau}{2} - \frac14\{H H^\dag, \circ\}
- \big[A_z,[A_{\bar{z}},\circ]\big] - \big[A_{\bar{z}},[A_z,\circ]\big],
\label{eq:K55}
\end{align}
acting on 
\beq \eta \equiv
\begin{pmatrix}
\dH \\
\dH^\dag\\
\dA_{\bar{z}}\\
\dA_{z}\\
\delta\phi
\end{pmatrix}, \eeq
where the rows of the operator correspond to the fluctuations of the 
self-dual equation \eqref{eq:YMCS-linearfluct1}, its Hermitian
conjugate, the fluctuations of the magnetic flux equation
\eqref{eq:YMCS-linearfluct2}, the gauge fixing condition
\eqref{eq:gauge_fixing_cond} and the fluctuations of the adjoint
scalar field equation \eqref{eq:YMCS-linearfluct3},
and $\circ$ is a placeholder for the matrix that the operator acts
on in $\eta$, which makes left and right actions more convenient in
matrix notation.
We have used the Coulomb gauge condition for the background gauge
field in eq.~\eqref{eq:K55}, i.e.~$D_{\{\bar{z}}A_{z\}}=0$.
The index of this operator will result in the number of real
zeromodes, because we have included the complex self-dual BPS equation
twice.

The definition of the split of the operator $\bbD$ into $D$ and $K$ is
that all derivative operators are placed in $D$ and the rest in $K$. 

The adjoint operator is found to be
\begin{align}
\mathbb{D}^\dag &= \i D^\dag + K^\dag, \label{eq:decomp_adjoint}\\
D^\dag &\equiv
\begin{pmatrix}
\p_z & 0 & 0 & 0 & 0 \\
0 & \p_{\bar{z}} & 0 & 0 & 0 \\
0 & 0 & -\frac{1}{\sqrt{2}}\p_{\bar{z}} & \frac{1}{\sqrt{2}}\p_{\bar{z}} & 0 \\
0 & 0 & \frac{1}{\sqrt{2}}\p_z & \frac{1}{\sqrt{2}}\p_z & 0 \\
0 & 0 & 0 & 0 & -\i 2D_{\bar{z}}\p_z -\i 2D_z \p_{\bar{z}} +\i2\p_{\bar{z}}\p_z
\end{pmatrix}
, \label{eq:Ddagop}\\
K^\dag &\equiv
\begin{pmatrix}
-A_z & 0 & -\frac{1}{4\sqrt{2}}\circ H & \circ X^\dag & K_{15}^\dag \\
0 & \circ A_{\bar{z}} & -\frac{1}{4\sqrt{2}}H^\dag & Y^\dag & K_{25}^\dag \\
-\circ H^\dag & 0 & \frac{1}{\sqrt{2}}[A_{\bar{z}},\circ] &
  -\frac{1}{\sqrt{2}}[A_{\bar{z}},\circ] & K_{35}^\dag \\
0 & H & -\frac{1}{\sqrt{2}}[A_z,\circ] & 
  -\frac{1}{\sqrt{2}}[A_z,\circ] & K_{45}^\dag \\
0 & 0 & \frac{1}{2\sqrt{2}} & 0 & K_{55}
\end{pmatrix},\\
K_{15}^\dag &\equiv \frac{\tau}{4}\circ H - \frac14\{\circ,\phi\}H
+\i[\circ,\phi]X^\dag,\label{eq:K15dag}\\
K_{25}^\dag &\equiv \frac{\tau}{4}H^\dag - \frac14H^\dag\{\circ,\phi\}
+\i Y^\dag[\circ,\phi],\label{eq:K25dag}\\
K_{35}^\dag &\equiv -\i 2[\circ,D_{\bar{z}}\phi],\label{eq:K35dag}\\
K_{45}^\dag &\equiv -\i 2[\circ,D_z\phi].\label{eq:K45dag}
\end{align}
First it is demonstrated that $D^\dag D=DD^\dag$ is the following
diagonal operator
\begin{align}
\Delta &= -D^\dag D = -DD^\dag  = 
\begin{pmatrix}
-\mathbf{1}_{N}\p_{\bar{z}}\p_z & 0 & 0 & 0 & 0 \\
0 & -\mathbf{1}_{N}\p_{\bar{z}}\p_z & 0 & 0 & 0 \\
0 & 0 & -\mathbf{1}_{N}\p_{\bar{z}}\p_z & 0 & 0 \\
0 & 0 & 0 & -\mathbf{1}_{N}\p_{\bar{z}}\p_z & 0 \\
0 & 0 & 0 & 0 & \square
\end{pmatrix}, \label{eq:Delta_def}\\
\square &\equiv
4(D_{\bar{z}}\p_z + D_z\p_{\bar{z}} - \p_{\bar{z}}\p_z)^2.
\end{align}

The index that we will calculate here is an extension of the index
that was calculated in ref.~\cite{Eto:2010mu}, which in turn is an
extension of the Atiyah-Singer index to $\mathbb{R}^2$, i.e.~of
Callias type, specialized to non-Abelian systems whose fluctuations
are described by $\mathbb{D}$. 
The index is formally given by
\beq \mathcal{I}
= \dim\kernel\mathbb{D} - \dim\kernel\mathbb{D}^\dag.
\eeq
If we can show that the dimension of the kernel of the adjoint
operator, $\mathbb{D}^\dag$, is zero, then the index is equal to the
number of zeromodes of the operator $\mathbb{D}$.
If not, the index only gives a lower bound on the number of zeromodes.

It will prove convenient to write the index as
\beq
\calI = \dim\kernel\bbD^\dag\bbD - \dim\kernel\bbD\bbD^\dag,
\eeq
which can be seen to hold based on the considerations
\beq
\langle\eta,\bbD^\dag\bbD\eta\rangle
=\langle\bbD\eta,\bbD\eta\rangle
=\|\bbD\eta\|^2\geq 0,
\eeq
and hence the kernel of $\bbD^\dag\bbD$ must coincide with that of
$\bbD$.
Similarly, the kernel of $\bbD\bbD^\dag$ must coincide with that of
$\bbD^\dag$. 
We will now write the index in the form
\begin{align}
\calI = \lim_{M\to\infty}
\calI(M^2) &= 
\lim_{M\to\infty}\Tr\left(\frac{\sfM^2}{\mathbb{D}^\dag\mathbb{D}+\sfM^2}\right)
-\Tr\left(\frac{\sfM^2}{\mathbb{D}\mathbb{D}^\dag+\sfM^2}\right)\non
&= \lim_{M\to\infty}\int_{\mathbb{C}}\tr\left\langle x\left|\left(
\frac{\sfM^2}{\mathbb{D}^\dag\mathbb{D}+\sfM^2}
-\frac{\sfM^2}{\mathbb{D}\mathbb{D}^\dag+\sfM^2}
\right)\right| x\right\rangle\;\d^2x, \label{eq:explicit_index}
\end{align}
where $\Tr$ denotes both a trace over states as well as the trace over
matrices, whereas $\tr$ is the usual matrix trace and we have defined
\beq
\sfM^2 \equiv {\textrm{block-diag}}\left[
  M^2\mathbf{1}_N, M^2\mathbf{1}_N, M^2\mathbf{1}_N, M^2\mathbf{1}_N, M^4\mathbf{1}_N
  \right].
\eeq
Notice that the last block contains $M^4$, matching the higher power
of derivatives in $\Delta$ (see eq.~\eqref{eq:Delta_def}). 

Subtleties arise in the case that the continuum part of the spectrum
extends down to zero and more care has to be taken.
This will not be a problem when the system possesses a mass gap, which
is the case in the fully Higgsed phase.

Using the decompositions \eqref{eq:decomp} and
\eqref{eq:decomp_adjoint}, we have
\begin{align}
\mathbb{D}^\dag\mathbb{D} = 
-D^\dag D + \i D^\dag K + \i K^\dag D + K^\dag K \equiv \Delta - L_1,
\\
\mathbb{D}\mathbb{D}^\dag =
-D D^\dag + \i D K^\dag + \i K D^\dag + K K^\dag \equiv \Delta - L_2,
\end{align}
and $L_{1,2}$ are at most linear in the operators $D$ and $D^\dag$. 
We will now follow ref.~\cite{Eto:2010mu} and expand
eq.~\eqref{eq:explicit_index} as follows
\begin{align}
  \frac{\sfM^2}{\bbD^\dag\bbD + \sfM^2} &=
  \sfM^2\left(P^{-1} + P^{-1} L_1P^{-1} + P^{-1} L_1P^{-1}L_1P^{-1} + \cdots\right),\label{eq:exp1}\\
  \frac{\sfM^2}{\bbD\bbD^\dag + \sfM^2} &=
  \sfM^2\left(P^{-1} + P^{-1} L_2P^{-1} + P^{-1} L_2P^{-1}L_2P^{-1} + \cdots\right),\label{eq:exp2}
\end{align}
where we have defined $P\equiv\Delta+\sfM^2$.
All terms beyond the second order in $L_{1,2}$ will not contribute to
the index \eqref{eq:explicit_index} in the $M\to\infty$ limit.
The symbolic manipulation of the terms given in equations
\eqref{eq:exp1} and \eqref{eq:exp2} follow analogously to that in
refs.~\cite{Lee:1991td,Eto:2010mu}, because we have that
\begin{equation}
  [P^{-1},D] = [P^{-1},D^\dag] = 0, \qquad
  D P^{-1} D^\dag = D^\dag P^{-1} D = -\mathbf{1} + \sfM^2 P^{-1} =
  -\mathbf{1} + P^{-1}\sfM^2.
\end{equation}
The result of the algebraic manipulations is \cite{Lee:1991td,Eto:2010mu}
\begin{align}
\mathcal{I} =
\lim_{M\to\infty} \i \int_{\mathbb{C}} \tr \sfM^2
\left\langle x \left| \frac{1}{\Delta+\sfM^2}
\big(D K^\dag - K^\dag D - D^\dag K + KD^\dag \big)
\frac{1}{\Delta+\sfM^2} \right| x\right\rangle\;\d^2x.
\label{eq:indexsimplified}
\end{align}
Calculating explicitly the matrix 
$D K^\dag - K^\dag D - D^\dag K + KD^\dag$, and using the fact
that\footnote{Here we adopt the notation that the indices $ij$ denote
the matrix block of the operator.}  
$K^\dag_{55} = K_{55}$, $K^\dag_{34} = K_{34}$, $K^\dag_{43} = K_{43}$
are self-adjoint operators, we find that the only contributions to the
trace are
\begin{align}
{\textrm{block-diag}}\left[
\p_{\bar{z}}K^\dag_{11} - \p_z K_{11},
\p_z K^\dag_{22} - \p_{\bar{z}}K_{22},
\frac{\p_{\bar{z}}K_{33} - \p_z K^\dag_{33}}{\sqrt{2}},
\frac{\p_{\bar{z}}K^\dag_{44} - \p_z K_{44}}{\sqrt{2}},
0\right].
\end{align}
Notice that the last block is identically zero, which has the
consequence that the $\square$ operator does not contribute to the
trace at all. This is due to the form of the operators $\bbD$,
$\bbD^\dag$ and the self-adjointness of $K_{55}$.
Furthermore, the matrix trace of the third and fourth block vanishes
\beq
\tr\big(\p_{\bar{z}}K_{33} - \p_z K^\dag_{33}\big) = 0, \qquad
\tr\big(\p_{\bar{z}}K^\dag_{44} - \p_z K_{44}\big) = 0,
\eeq
simply because the trace of a commutator vanishes.
Thus the index simplifies to
\begin{align}
\mathcal{I} &=
\lim_{M\to\infty} -2\i M^2\int_{\mathbb{C}}\tr
(\p_{\bar{z}} A_z - \p_z A_{\bar{z}})
\left\langle x \left|(-\p_{\bar{z}}\p_z+M^2)^{-2}
\right| x\right\rangle\;\d^2x\non
&= %
\lim_{M\to\infty} -M^2
\frac{N}{\sqrt{2N}}
\sum_{f=1}^{\NF}\int_{\mathbb{C}} F_{12}^0\;\d^2x
\int\frac{\d^2k}{\left(2\pi\right)^2}
\frac{1}{\left(\frac{1}{4}k^2+M^2\right)^{2}}\non
&= %
-\NF N\frac{1}{\pi\sqrt{2N}}\int_{\mathbb{C}} F_{12}^0\;\d^2x \non
&= %
2\NF N\nu = \frac{2\NF N k}{n_0}.
\label{eq:index}
\end{align}
Now, recall that the operator $\bbD$ contained both the self-dual
equation as well as its Hermitian conjugate equation, 
$\delta(D_{\bar{z}}H)^\dag$, and therefore the index is counting the
number of real zeromodes.

Notice that leaving the gauge group $G'$ unspecified, the calculation
remains unaltered and depends only on $F_{12}^0$, although the
operator $\bbD$ now contains projections $\langle\circ\rangle_{G'}$.
Carrying out the index calculation for generic $G'$ thus gives the
result \eqref{eq:index} and the $G'$ dependence is simply encoded in
the greatest common divisor of the Abelian charges of the $G'$
invariants, which is exactly $n_0$.
For $G'=\SU(N)$, the index is $2\NF k$ and thus independent of $N$.
However, we assumed a fully Higgsed theory, which requires
$\NF\geq N$.
For the square-matrix case, $\NF=N$, the index is simply $2Nk$. 

In order to establish that $\calI$ is in fact the number of real
zeromodes, we first need the following theorem.

\begin{theorem}
  The kernel of $\bbD^\dag$, in the case of $G=\U(N)$ and $\NF\geq N$,
  has dimension zero.
  \label{thm:1}
\end{theorem}
\emph{Proof}:
First we act on a state with the adjoint operator $\bbD^\dag$:
\begin{align}
\mathbb{D}^\dag \begin{pmatrix} \alpha\\ \alpha^\dag\\ \beta\\ 
\beta^\dag\\ \omega \end{pmatrix}
=
\begin{pmatrix}
\i D_z\alpha - \frac{1}{4\sqrt{2}}\beta H + \beta^\dag X^\dag
+\frac{\tau}{4}\omega H - \frac14\{\omega,\phi\} H + \i[\omega,\phi]X^\dag\\
\i D_{\bar{z}}\alpha^\dag - \frac{1}{4\sqrt{2}}H^\dag\beta + Y^\dag\beta^\dag
+\frac{\tau}{4}H^\dag\omega - \frac14H^\dag\{\omega,\phi\} +\i Y^\dag[\omega,\phi]\\
-\alpha H^\dag - \frac{\i}{\sqrt{2}}D_{\bar{z}}\beta
+\frac{\i}{\sqrt{2}}D_{\bar{z}}\beta^\dag
-\i 2[\omega,D_{\bar{z}}\phi] \\
H \alpha^\dag + \frac{\i}{\sqrt{2}}D_z\beta
+\frac{\i}{\sqrt{2}}D_z\beta^\dag
-\i 2[\omega,D_z\phi] \\
D_{\bar{z}}D_z\omega + D_z D_{\bar{z}}\omega
+\frac{1}{2\sqrt{2}}\beta - \frac{\tau}{2}\omega
-\frac14\{H H^\dag,\omega\}
\end{pmatrix} = 0, 
\label{eq:Ddagvector}
\end{align}
where $\omega=\omega^\dag$.
Writing combinations of the above equations yields
\begin{align}
D_z\alpha &= 0, \label{eq:vanishymcs1}\\
\left(\frac{1}{\sqrt{2}}(\beta + \beta^\dag) - \tau\omega
  + \{\omega,\phi\} - \frac{\i}{\sqrt{2}}[\omega,\phi]\right)H &= 0, \label{eq:vanishymcs2}\\
D_{\bar{z}}\beta^\dag + \i\sqrt{2}\alpha H^\dag &= 0, \label{eq:vanishymcs3}\\
D_{\bar{z}}\beta + 2\sqrt{2} [\omega,D_{\bar{z}}\phi] &= 0,  \label{eq:vanishymcs4}\\ 
D_{\bar{z}}D_z\omega + D_z D_{\bar{z}}\omega + \frac{1}{2\sqrt{2}}\beta
  - \frac{\tau}{2}\omega - \frac14\{H H^\dag,\omega\} &= 0,
  \label{eq:vanishymcs5}
\end{align}
where we have chosen the gauge fixing functions as
\beq
X^\dag = Y = -\frac{1}{4\sqrt{2}}H.
\eeq
Writing the complex norm-squared of eq.~\eqref{eq:vanishymcs3}, we
have
\begin{align}
0 &= \int_{\mathbb{C}}\big(D_{\bar{z}}\beta^\dag + \i\sqrt{2}\alpha H^\dag\big)^\dag
\big(D_{\bar{z}}\beta^\dag + \i\sqrt{2}\alpha H^\dag\big)\;\d^2x \non
&=\|D_{\bar{z}}\beta^\dag\|^2
+2\|\alpha H^\dag\|^2
-\i\sqrt{2}\int_{\mathbb{C}}\p_{\bar{z}}(H\alpha^\dag\beta^\dag)\;\d^2x
+\i\sqrt{2}\int_{\mathbb{C}}\p_z(\beta\alpha H^\dag)\;\d^2x\non
&=\|D_{\bar{z}}\beta^\dag\|^2
+2\|\alpha H^\dag\|^2,
\label{eq:alpha_vanishes}
\end{align}
where in the second line, we have used the self-dual equation
$D_{\bar{z}}H=0$, eq.~\eqref{eq:vanishymcs1} and their Hermitian
conjugates, and we have defined
\beq
\|X\|^2 \equiv \int_{\mathbb{C}} X^\dag X\;\d^2x.
\eeq
In the third line, we have used that the boundary conditions for the
fluctuations are
\beq
\lim_{|z|\to\infty}\alpha
=\lim_{|z|\to\infty}\beta
=\lim_{|z|\to\infty}\omega = 0.
\label{eq:fluc_BC}
\eeq
From the third line in eq.~\eqref{eq:alpha_vanishes}, we can conclude
that
\beq
\alpha=0,
\eeq
since $H$ has full rank almost everywhere (except at vortex
positions) and that $D_{\bar{z}}\beta^\dag=0$.
Taking the Hermitian conjugate of eq.~\eqref{eq:vanishymcs5}, we can
conclude that
\beq
\beta=\beta^\dag,
\eeq
is a real-valued adjoint scalar field and therefore we have
\beq
D_{\bar{z}}\beta^\dag = D_{\bar{z}}\beta = D_z\beta^\dag = D_z\beta = 0.
\eeq
From this result, we have from eq.~\eqref{eq:vanishymcs4} that
\beq
[\omega,D_{\bar{z}}\phi]=0,
\eeq
which is it not quite strong enough a condition to conclude that
$\omega$ vanishes.
Considering first $D_{\bar{z}}\beta=0$ in the vortex background, using
the solution to the gauge field \eqref{eq:mm}, we have
\beq
\p_{\bar{z}}\beta + [S(z,\bar{z})^{-1}\p_{\bar{z}}S(z,\bar{z}),\beta]
= 0,
\eeq
which yields the solution
\beq
\beta = S(z,\bar{z})^{-1}\beta_0(z)S(z,\bar{z}).
\eeq
Considering instead $D_z\beta=0$ in the vortex background, using the
Hermitian conjugate of the solution to the gauge field
\eqref{eq:mm}, we obtain
\beq
\p_z\beta - [\p_z S(z,\bar{z})^\dag S(z,\bar{z})^{\dag-1},\beta]=0,
\eeq
yielding the different solution
\beq
\beta = S(z,\bar{z})^\dag\tilde\beta_0(\bar{z})S(z,\bar{z})^{\dag-1}.
\eeq
Equating the two solutions, we have
\beq
\tilde\beta_0(\bar{z}) = \Omega^{-1}\beta_0(z)\Omega,
\eeq
where $\Omega$ is the Hermitian invertible matrix defined in
eq.~\eqref{eq:mm_gauge_inv}.
There is no Hermitian matrix that can transform a holomorphic function
in $z$ into an anti-holomorphic function.
We can thus conclude that $\beta_0(z)$ and $\tilde\beta_0(\bar{z})$
must be independent of $z$ and $\bar{z}$ and hence be constant matrices.
That is not sufficient, however, for meeting the constraint that
$\beta=\beta^\dag$ and hence, we must have that
$S(z,\bar{z})^{-1}=S(z,\bar{z})^\dag$ and $\beta_0=\beta_0^\dag$ (and
$\tilde\beta_0=\tilde\beta_0^\dag$), which is generically not the case
for a background solution.
If $S(z,\bar{z})^{-1}\neq S(z,\bar{z})^\dag$, we must take
\beq
\beta_0 = \tilde\beta_0 = c\mathbf{1}_N,
\eeq
which means that
\beq
\beta = c\mathbf{1}_N,
\eeq
or else we can have
\beq
\beta = S(z,\bar{z})\beta_0 S(z,\bar{z})^{-1}
= S(z,\bar{z})\beta_0 S(z,\bar{z})^\dag,
\eeq
with $\beta_0=\beta_0^\dag$ a constant matrix.
Using the boundary conditions \eqref{eq:fluc_BC}, we have $c=0$ or
$\beta_0=0$ and hence $\beta=0$ in any case.
This can be seen by the fact that $S(z,\bar{z})$ in the latter case
acts as a unitary transformation of a Hermitian constant matrix that
must have real eigenvalues: the only way the matrix can
vanish at spatial infinity is if all the eigenvalues vanish.

Since $\beta=0$ and $H$ has full rank, almost everywhere (except at
vortex positions), we conclude from eq.~\eqref{eq:vanishymcs2} that
$\omega=0$. 
This can be seen by choosing generic values of $\tau$ for which
cancellations are certainly impossible.
Thus all fluctuations $\alpha=\beta=\omega=0$ vanish which completes the
proof.
\hfill$\square$

The extension to unequal Abelian and non-Abelian couplings $e,\kappa$ 
and $g,\mu$ is straightforward and theorem \ref{thm:1} still holds.
The extension to generic semi-simple gauge groups $G'$ alters the
operators $\bbD$ and $\bbD^\dag$ in a straightforward fashion and
various places the projection operator $\langle\circ\rangle_{G}$
appears, but we can establish
\begin{theorem}
  The kernel of $\bbD^\dag$, in the case of generic gauge groups
  $(\U(1)\times G')/\mathbb{Z}_{n_0}$ with $G'$ a semi-simple group
  and $\NF\geq N$, has dimension zero. 
  \label{thm:2}
\end{theorem}
\emph{Proof}:
The proof is analogous to that of theorem \ref{thm:1}, with the
following changes.
For simplicity, we will provide the proof for the case with the
Abelian couplings equal to the non-Abelian ones.
First of all, the $\mathfrak{g}$-valued equations (where
$\mathfrak{g}$ is the algebra corresponding to the gauge group $G$)
now contain a projection of the Higgs fields to the group $G$ by means
of $\langle\circ\rangle_G$.
This amounts to the $K$ matrix of the form
\begin{align}
K &\equiv
\begin{pmatrix}
-A_{\bar{z}} & 0 & -\circ H & 0 & 0 \\
0 & \circ A_z & 0 & H^\dag & 0 \\
-\frac{1}{4\sqrt{2}} \circ H^\dag & -\frac{1}{4\sqrt{2}}H & 
  \frac{1}{\sqrt{2}}\left[A_z,\circ\right] & 
  -\frac{1}{\sqrt{2}}[A_{\bar{z}},\circ] & \frac{1}{2\sqrt{2}} \\
2\langle\circ X\rangle_G & 2\langle Y\circ\rangle_G & -\frac{1}{\sqrt{2}}[A_z,\circ] & 
  -\frac{1}{\sqrt{2}}[A_{\bar{z}},\circ] & 0 \\ 
K_{51} & K_{52} & K_{53} & K_{54} & K_{55}
\end{pmatrix},\\
K_{51} &\equiv \frac{\tau}{2}\langle\circ H^\dag\rangle_G
-\frac12\langle\{\circ H^\dag,\phi\}\rangle_G
-\i2[\langle\circ X\rangle_G,\phi],\\
K_{52} &\equiv \frac{\tau}{2}\langle H\circ\rangle_G
-\frac12\langle\{H \circ,\phi\}\rangle_G
-\i2[\langle Y\circ\rangle_G,\phi],\\
K_{55} &\equiv -\frac{\tau}{2} - \frac12\langle\{H H^\dag, \circ\}\rangle_G
- \big[A_z,[A_{\bar{z}},\circ]\big] - \big[A_{\bar{z}},[A_z,\circ]\big],
\end{align}
with $D$, $K_{53}$ and $K_{54}$ unchanged and given in
eqs.~\eqref{eq:Dop}, \eqref{eq:K53} and \eqref{eq:K54}, respectively.
The adjoint operator is found straightforwardly and we have
\begin{align}
K^\dag &\equiv
\begin{pmatrix}
-A_z & 0 & -\frac{1}{4\sqrt{2}}\circ H & \circ X^\dag & K_{15}^\dag \\
0 & \circ A_{\bar{z}} & -\frac{1}{4\sqrt{2}}H^\dag & Y^\dag & K_{25}^\dag \\
-2\langle\circ H^\dag\rangle_G & 0 & \frac{1}{\sqrt{2}}[A_{\bar{z}},\circ] &
  -\frac{1}{\sqrt{2}}[A_{\bar{z}},\circ] & K_{35}^\dag \\
0 & 2\langle H\circ\rangle_G & -\frac{1}{\sqrt{2}}[A_z,\circ] & 
  -\frac{1}{\sqrt{2}}[A_z,\circ] & K_{45}^\dag \\
0 & 0 & \frac{1}{2\sqrt{2}} & 0 & K_{55}
\end{pmatrix},
\end{align}
with $D^\dag$, $K_{15}^\dag$, $K_{25}^\dag$, $K_{35}^\dag$,
$K_{45}^\dag$ and $K_{55}$ unchanged and given in
eqs.~\eqref{eq:Ddagop}, \eqref{eq:K15dag}-\eqref{eq:K45dag}, and
\eqref{eq:K55}, respectively.

Acting on $\eta$ with $\bbD^\dag$ corresponding to the general gauge
group case at hand, we obtain that the eqs.~\eqref{eq:vanishymcs1},
\eqref{eq:vanishymcs2} and \eqref{eq:vanishymcs4} remain unaltered,
whereas eq.~\eqref{eq:vanishymcs3} changes into
\beq
D_{\bar{z}}\beta^\dag + \i2\sqrt{2}\langle\alpha H^\dag\rangle_G &= 0, \label{eq:vanishymcs3gen}
\eeq
and eq.~\eqref{eq:vanishymcs5} remains Hermitian apart from $\beta$.
Eq.~\eqref{eq:alpha_vanishes}, in the general gauge group case, can be
written as
\begin{align}
0 &= \tr\int_{\mathbb{C}}\big(D_{\bar{z}}\beta^\dag + \i 2\sqrt{2}\langle\alpha H^\dag\rangle_G\big)^\dag
\big(D_{\bar{z}}\beta^\dag + \i 2\sqrt{2}\langle\alpha H^\dag\rangle_G\big)\;\d^2x \non
&=\tr\|D_{\bar{z}}\beta^\dag\|^2
+8\tr\|\langle\alpha H^\dag\rangle_G\|^2
-\i\sqrt{2}\tr\int_{\mathbb{C}}\p_{\bar{z}}(H\alpha^\dag\beta^\dag)\;\d^2x
+\i\sqrt{2}\tr\int_{\mathbb{C}}\p_z(\beta\alpha H^\dag)\;\d^2x\non
&=\tr\|D_{\bar{z}}\beta^\dag\|^2
+8\tr\|\langle\alpha H^\dag\rangle_G\|^2.
\label{eq:alpha_vanishes_gen}
\end{align}
We thus get $D_{\bar{z}}\beta^\dag=0$ as before, but now we have
$\langle\alpha H^\dag\rangle_G=0$, which is exactly the condition
found in ref.~\cite{Eto:2009bg}.
In the latter reference, it was shown that $\alpha$ must vanish if
there are no independent $G'$ invariants with positive $\U(1)$
windings.
This is the case for $G'=\SU(N)$, $G'=\SO(N)$ and $G'=\USp(2M)$.

Replaying now the logic of the proof of theorem \ref{thm:1}, we have
$\alpha=0$ and $D_{\bar{z}}\beta^\dag=0$ as before.
Taking the Hermitian conjugate of eq.~\eqref{eq:vanishymcs5}, which
now contains some projection operators to $G$, reveals that
$\beta=\beta^\dag$ and hence we have $D_{\bar{z}}\beta=D_z\beta=0$ as
before.
Using the remaining arguments of the proof of theorem \ref{thm:1}, we
thus again reach the conclusion that $\alpha=\beta=\omega=0$.
\hfill$\square$

\begin{corollary}
The operator $\bbD^\dag$ does not contain zeromodes as shown by
theorems \ref{thm:1} and \ref{thm:2}, and therefore the index
\eqref{eq:index} counts the number of real zeromodes of 
$\bbD$ and therefore of the BPS equations \eqref{eq:solvable},
\eqref{eq:master1}, and \eqref{eq:master3}.
The number of zeromodes is the dimension of the moduli space,
which is
\beq
\dim_{\mathbb{C}}\mathcal{M}_{k,N,\NF} = \frac{k N \NF}{n_0}.
\eeq
\end{corollary}


\section{The moduli matrix method}\label{sec:modulimatrix}

The BPS equations obtained in sec.~\ref{sec:model} have many solutions
with degenerate energy. The space of solutions forms a manifold which
is the so-called the moduli space\footnote{By moduli space, we here
  mean the moduli space of solutions and not the moduli space of vacua
  which is the manifold of the vacua in the theory. }. 
We will now solve the BPS equations and write down the master
equations for the YMCSH vortices of
vorticity $k$.

The moduli matrix approach solves the first BPS equation
\eqref{eq:solvable}:
\beq
H = {S}^{-1}H_0(z),\quad 
A_{\bar{z}} = -\i S^{-1}\p_{\bar{z}}S,\qquad
S \in G^{\mathbb{C}}.
\label{eq:mm}
\eeq
The gauge field $A_{\bar{z}}=(A_1+\i A_2)/2$ enjoys the complexified
the gauge symmetry $G^{\mathbb{C}}$, which for $G=\U(N)$ is simply
$\GL(N,\mathbb{C})$.
The gauge symmetry and the flavor symmetry act on the new matrices
with a left and right action, respectively,
\beq
S^{-1} \to U_{\rm c} S^{-1},\qquad
H_0 \to H_0 U_{\rm f},
\eeq
where $U_{\rm c}\in G^{\mathbb{C}}$ and $U_{\rm f}\in\SU(\NF)$.
Furthermore, there is a residual gauge symmetry, which is
denoted $V$-equivalence and it acts as
\beq
S \sim V(z) S,\qquad
H_0 \sim V(z) H_0,\qquad
V(z) \in G^{\mathbb{C}},
\label{eq:v-equiv}
\eeq
with $V(z)$ being holomorphic with respect to the coordinate $z$.
We are now left with the remaining four BPS equations
\eqref{eq:master1}, \eqref{eq:master3}.
In order to rewrite these equations in a gauge invariant fashion, let
us introduce the following gauge invariants:
\beq
\Omega = SS^\dagger, \qquad 
\Omega_0 = v^{-2} H_0H_0^\dagger, \qquad 
\Upsilon  = S\phi S^{-1}.
\label{eq:mm_gauge_inv}
\eeq
In terms of these invariants, the field strength reads
\beq
F_{12} &=& -2[D_z,D_{\bar{z}}] = -2 S^{-1} \p_{\bar{z}}(\p_z \Omega \Omega^{-1}) S, \\
F_{0\bar z} &=& - \i[D_0,D_{\bar{z}}] = \frac{F_{01}+\i F_{02}}{2} 
= -D_{\bar{z}}\phi
= -S^{-1} \p_{\bar{z}} \Upsilon S,
\eeq
where we have used $D_0H=\i\phi H$ and that we are considering
only the static case $\p_0=0$. 
The complex covariant Laplacian on $\phi$ is
\beq
D_z D_{\bar{z}} \phi = \i D_z[D_0,D_{\bar{z}}] 
= S^{-1} \left(\p_z\p_{\bar{z}} \Upsilon + 
[\p_{\bar{z}}\Upsilon,\p_z\Omega \Omega^{-1}]\right)S.
\eeq
We can now rewrite eqs.~\eqref{eq:master1} and \eqref{eq:master3} using
the gauge invariants. To this end, let us decompose the Abelian and
non-Abelian parts as $S=s\hat S$ with 
$s\in\U(1)^{\mathbb{C}}$ and $\hat S\in {G'}^{\mathbb{C}}$:
\beq
\Omega = \omega \hat \Omega, \quad 
\omega = |s|^2, \quad 
\hat{\Omega} = \hat S \hat S^\dagger, \quad
\Upsilon = \Upsilon^0 t^0 + \hat \Upsilon, \quad 
\Upsilon^0 = \phi^0, \quad
\hat{\Upsilon} = \hat S \hat \phi \hat S^{-1}.
\eeq
This leads to the system of master equations 
\begin{align}
\p_{\bar{z}}(\p_z \hat \Omega{\hat \Omega}^{-1}) &=
- \frac{m_g^2}{2}\langle\Omega_0\Omega^{-1}\rangle_{G'}
+ \frac{m_\mu}{2}\hat{\Upsilon}, 
\label{eq:me1}\\
\p_{\bar{z}}\p_z\log\omega &= 
\frac{m_e^2}{4N}\tr[{\bf 1}_N-
\Omega_0{\Omega}^{-1}]
+\frac{m_\kappa}{2\sqrt{2N}}\Upsilon^0, 
\label{eq:me2}\\
\p_z\p_{\bar{z}}\hat\Upsilon
+[\p_{\bar{z}}\hat\Upsilon,\p_z\hat\Omega\hat\Omega^{-1}] &=
\frac{m_\mu^2}{4} \hat\Upsilon
-\frac{m_g^2m_\mu}{4}\langle\Omega_0\Omega^{-1}\rangle_{G'}
-\frac{m_g^2}{4}\big[\langle\Omega_0\Omega^{-1}\rangle_{G'},\hat\Upsilon\big]\non
&\phantom{=\ }
+\frac{m_g^2}{4}\big\langle\{\Omega_0\Omega^{-1},\Upsilon\}\big\rangle_{G'},
\label{eq:me3}\\
\p_z\p_{\bar{z}} \Upsilon^0 &= 
\frac{1}{4\sqrt{2N}}\tr\left[
2( m_\kappa^2 + m_e^2 \Omega_0\Omega^{-1}) \Upsilon + {m_e^2m_\kappa}
({\bf 1}_N - \Omega_0\Omega^{-1})
\right],
\label{eq:me4}
\end{align}
where, for convenience, we recall the definitions of the mass
parameters of the theory
\begin{align}
m_g \equiv g v, \qquad
m_e \equiv e v, \qquad
m_\mu \equiv \frac{\mu g^2}{4\pi}, \qquad
m_\kappa \equiv \frac{\kappa e^2}{4\pi},
\end{align}
and the boundary conditions for the PDEs are given by
\beq
\lim_{|z|\to\infty} \Omega = \Omega_0, \qquad
\lim_{|z|\to\infty} \Upsilon = 0.
\eeq
A first simplification of the equations happens when considering the
equal coupling case, $e=g$ and $\kappa=\mu$, for which we can
write
\begin{align}
\p_{\bar{z}}(\p_z\Omega\Omega^{-1}) &=
\frac{m_g^2}{2}\langle\mathbf{1}_N - \Omega_0\Omega^{-1}\rangle_{G}
+\frac{m_\mu}{2}\Upsilon,\\
\p_z\p_{\bar{z}}\Upsilon
+[\p_{\bar{z}}\Upsilon,\p_z\Omega\Omega^{-1}] &=
\frac{m_\mu^2}{4}\Upsilon
+\frac{m_g^2m_\mu}{4}\langle\mathbf{1}_N - \Omega_0\Omega^{-1}\rangle_{G}
-\frac{m_g^2}{4}\big[\langle\Omega_0\Omega^{-1}\rangle_{G},\Upsilon\big]\non
&\phantom{=\ }
+\frac{m_g^2}{4}\big\langle\{\Omega_0\Omega^{-1},\Upsilon\}\big\rangle_{G}.
\end{align}
A further simplification happens when $G'=\SU(N)$, so the $G'$
projections simply give (a half times) the traceless part of the
matrix expressions.
This corresponds to the case of $G=\U(N)$ (due to the equal 
couplings):
\begin{align}
\p_{\bar{z}}(\p_z\Omega\Omega^{-1}) &=
\frac{m_g^2}{4}(\mathbf{1}_N - \Omega_0\Omega^{-1})
+\frac{m_\mu}{2}\Upsilon,\\
\p_z\p_{\bar{z}}\Upsilon
+[\p_{\bar{z}}\Upsilon,\p_z\Omega\Omega^{-1}] &=
\frac{m_\mu^2}{4}\Upsilon
+\frac{m_g^2m_\mu}{8}(\mathbf{1}_N - \Omega_0\Omega^{-1})
+\frac{m_g^2}{4}\Upsilon\Omega_0\Omega^{-1}.
\end{align}
The zeromodes or moduli are all contained in the moduli matrix
$H_0(z)$, which appears in the combination
$\Omega_0=v^{-2}H_0H_0^\dag$ in the master equations above.
Although we have not proved from the PDE point-of-view that the field
$\Omega$ and $\Upsilon$ are uniquely determined by the above PDEs for
given $H_0(z)$, we know that the number of moduli parameters available
in the moduli matrix are (see
refs.~\cite{Eto:2005yh,Eto:2006pg,Eto:2008yi,Eto:2009bg,Gudnason:2009ut,Eto:2010mu}) 
\beq
\dim_{\mathbb{C}}\mathcal{M}_{k,N,\NF} = \frac{k N \NF}{n_0},
\eeq
which is in perfect agreement with the index theorem of
sec.~\ref{sec:indextheorem}.
For explicit realizations of the moduli in terms of the moduli
matrices $H_0(z)$ for $G'=\SU(N)$, see
refs.~\cite{Eto:2005yh,Eto:2006pg}, whereas for $G'=\SO(N)$ and
$G'=\USp(2M)$, see refs.~\cite{Eto:2009bg,Eto:2010mu}.

\subsection{Examples}

The master equations are in general quite complicated due to their
non-Abelian nature.
A nice simplification happens when considering the centers of a patch
of the moduli space, for which all matrices may be taken to be
diagonal matrices.
For simplicity, we will take the $G=\U(N)$ case, for which we have
\begin{align}
  \nabla^2 u_i &= \frac{m_g^2}{2}\big(e^{2u_i} - 1\big) - m_\mu\Upsilon_i
  +2\pi\sum_{r=1}^{k_i}\delta^{(2)}(z-Z_r^i), \qquad i=1,2,\ldots,N,\label{eq:ex_ui}\\
  \nabla^2\Upsilon_i &= m_\mu^2\Upsilon_i
  - \frac{m_g^2m_\mu}{2}\big(e^{2u_i} - 1\big) + m_g^2\Upsilon_i e^{2u_i},\label{eq:ex_Upsiloni}
\end{align}
where $i$ is nowhere summed over, we have defined the field
\beq
u_i \equiv \frac12\log(\Omega_0\Omega^{-1})_i,
\label{eq:u_def}
\eeq
the subscript (index) $i$ denotes here the $i$-th diagonal element
of each diagonal matrix, and the vortex positions are encoded in the
position moduli $Z_r^i$.
Notice that only the equation for $u_i$ contains delta function
sources for the vortices.
Notice also that there are no orientational moduli in this example,
because considering the center of a patch of the moduli space simply
means setting the orientational moduli to zero in some local
coordinates.
For analysis or numerical computations, it is convenient to perform
the rescaling \eqref{eq:rescaling}, yielding
\begin{align}
  \nabla^2 u_i &= \frac12\big(e^{2u_i} - 1\big) - \Upsilon_i
  +2\pi\sum_{r=1}^{k_i}\delta^{(2)}(z-Z_r^i), \qquad i=1,2,\ldots,N,\\
  \nabla^2\Upsilon_i &= \tau\Upsilon_i
  - \frac{\tau}{2}\big(e^{2u_i} - 1\big) + \Upsilon_i e^{2u_i},
\end{align}
where $\tau\geq 0$ is defined in eq.~\eqref{eq:tau}.

An example of vortex equations in the case of unequal couplings
in the $G'=\SU(2)$ theory (for simplicity) is
\begin{align}
  \nabla^2 u_1 &= \frac{m_e^2}{4}\big(e^{2u_1} + e^{2u_2} - 2\big)
  +\frac{m_g^2}{4}\big(e^{2u_1} - e^{2u_2}\big)
  -\frac{m_\kappa}{2}\Upsilon^0 - \frac{m_\mu}{2}\Upsilon^3
  +2\pi\sum_{r=1}^{k_1}\delta^{(2)}(z-Z_r^1),\non
  \nabla^2 u_2 &= \frac{m_e^2}{4}\big(e^{2u_1} + e^{2u_2} - 2\big)
  -\frac{m_g^2}{4}\big(e^{2u_1} - e^{2u_2}\big)
  -\frac{m_\kappa}{2}\Upsilon^0 + \frac{m_\mu}{2}\Upsilon^3
  +2\pi\sum_{r=1}^{k_2}\delta^{(2)}(z-Z_r^2),\non
  \nabla^2\Upsilon^0 &= m_\kappa^2\Upsilon^0
  +\frac{m_e^2}{2}\big(e^{2u_1} + e^{2u_2}\big)(\Upsilon^0 - m_\kappa)
  +\frac{m_e^2}{2}\big(e^{2u_1} - e^{2u_2}\big)\Upsilon^3 + m_e^2m_\kappa,\non
  \nabla^2\Upsilon^3 &= m_\mu^2\Upsilon^3
  +\frac{m_g^2}{2}\big(e^{2u_1} - e^{2u_2}\big)(\Upsilon^0 - m_\mu)
  +\frac{m_g^2}{2}\big(e^{2u_1} + e^{2u_2}\big)\Upsilon^3,
\end{align}
where $u_{1,2}$ are given by eq.~\eqref{eq:u_def} and
$\Upsilon=\frac12\Upsilon^0\mathbf{1}_2+\frac12\Upsilon^3\sigma^3$
with $\sigma^3$ being the third Pauli spin matrix.
Notice that this example correctly reduces to
eqs.~\eqref{eq:ex_ui}-\eqref{eq:ex_Upsiloni} in the equal 
coupling case ($m_g=m_e$ and $m_\mu=m_\kappa$).
Performing a rescaling
\beq
\p_z\to m_g^2\p_z, \qquad
\Upsilon^0\to \frac{m_g^2}{m_\kappa}\Upsilon^0, \qquad
\Upsilon^3\to \frac{m_g^2}{m_\mu}\Upsilon^3,
\eeq
convenient for further analysis, we get
\begin{align}
  \nabla^2 u_1 &= \frac{\gamma_{e/g}}{4}\big(e^{2u_1} + e^{2u_2} - 2\big)
  +\frac14\big(e^{2u_1} - e^{2u_2}\big)
  -\frac12\Upsilon^0 - \frac12\Upsilon^3
  +2\pi\sum_{r=1}^{k_1}\delta^{(2)}(z-Z_r^1),\non
  \nabla^2 u_2 &= \frac{\gamma_{e/g}}{4}\big(e^{2u_1} + e^{2u_2} - 2\big)
  -\frac14\big(e^{2u_1} - e^{2u_2}\big)
  -\frac12\Upsilon^0 + \frac12\Upsilon^3
  +2\pi\sum_{r=1}^{k_2}\delta^{(2)}(z-Z_r^2),\non
  \nabla^2\Upsilon^0 &= \gamma_{\kappa/g}\Upsilon^0
  +\frac{\gamma_{e/g}}{2}\big(e^{2u_1} + e^{2u_2}\big)(\Upsilon^0 - \gamma_{\kappa/g})
  +\frac{\gamma_{e/g}}{2}\big(e^{2u_1} - e^{2u_2}\big)\Upsilon^3 + \gamma_{e/g}\gamma_{\kappa/g},\non
  \nabla^2\Upsilon^3 &= \gamma_{\mu/g}\Upsilon^3
  +\frac12\big(e^{2u_1} - e^{2u_2}\big)(\Upsilon^0 - \gamma_{\mu/g})
  +\frac12\big(e^{2u_1} + e^{2u_2}\big)\Upsilon^3,
\end{align}
where $\gamma_{i/j}$ are defined in eq.~\eqref{eq:gammas}.
Notice that there are exactly three dimensionless couplings:
$\gamma_{e/g}$, $\gamma_{\kappa/g}$ and $\gamma_{\mu/g}$, which
parametrize the most general coupling space of the theory.

\section{The D-brane picture}\label{sec:dbranepicture}

We will now study the YMCSH vortices from a
D-brane perspective with which we can geometrically understand some
properties of solitons in a low energy effective theory of the D-brane
configuration. 
We will show that the results in the previous sections can be
reproduced by making use of the language in Type IIB string theory
which is simple, intuitive and geometrical.
Our construction here is in some sense a non-Abelianization of the
construction made in ref.~\cite{Ohta:1999gj}, where Abelian
Chern-Simons-Higgs vortices were constructed in Type IIB string
theory. 

The Type IIB D-brane configuration of interest is summarized in
the table below. 
\begin{center}
\begin{tabular}{c|cccccccccc}
Type IIB & 1 & 2 & 3 & 4 & 5 & 6 & 7 & 8 & 9\\
\hline
NS5 & $\heartsuit$ & $\heartsuit$ & $\heartsuit$ & $\heartsuit$ & $\heartsuit$ & -- & -- & -- & --\\
($\kappa$,1)5 & $\heartsuit$ & $\heartsuit$ & $\heartsuit_{\cos\theta}$ & -- 
& -- & -- & $\heartsuit_{\sin\theta}$ & $\heartsuit$ & $\heartsuit$\\
$\NC$ D3 & $\heartsuit$ & $\heartsuit$ & -- & -- & -- & $|\heartsuit|$ & -- & -- & -- \\
$\NF$ D5 & $\heartsuit$ & $\heartsuit$ & -- & -- & -- & -- & $\heartsuit$ & $\heartsuit$ & $\heartsuit$
\end{tabular}
\end{center}
The $(\kappa,1)$5-brane which is the bound state of an NS5-brane
(spanned in the directions 012389) and $\kappa$ D5-brane (spanned in
the directions 012789) is oriented at an angle
$\theta$ in $(x^3,x^7)$-space. 
The D3-brane is suspended by the ``left" NS5-brane at $x^6=0$ and the 
``right" $(\kappa,1)$5-brane at $x^6 = L_6$.

A low energy effective theory on $\NC$ D3-branes suspended by the five
branes is a $d=2+1$ ${\cal N}=2$ supersymmetric $U(\NC)$ YMCS theory
in a limit where massive string modes of order $l_s^{-1}$ ($l_s$ is
the string scale) are decoupled. 
The $U(\NC)$ gauge coupling is given by
\beq
\frac{1}{g^2} = \frac{L_6}{2\pi g_s},
\label{eq:gc}
\eeq
with $g_s$ being the string coupling constant of Type IIB string theory. 
Note that $g_s$ is a dimensionless coupling constant, so $g^2$ is
dimensionful in 2+1 dimension, as it should be.

The Chern-Simons (CS) interaction appears in the low-energy effective 
theory because of the D3-branes suspended between an NS5-brane and a 
$(p,q)$5-brane \cite{Kitao:1998mf}. 
Let us briefly review how the CS term appears \cite{Kitao:1998mf}.
To this end, we take the simple example where a D3-brane is suspended
by the 5-branes. 
From the supergravity (SUGRA) solution of the $(p,q)$5-brane, there is
a non-trivial VEV of the axion field \cite{Schwarz:1995dk}
\beq
\chi(x^4,x^5,x^6,x^7)
= \frac{1}{g_s} \frac{\sin\theta\cos\theta(1-H)}{\sin^2\theta + H \cos^2\theta},
\eeq
with a harmonic function $H=1+l_s^2/r^2$ on the
$(x^4,x^5,x^6,x^7)$-space ($r$ is the distance from the
$(p,q)$5-brane). Clearly, the axion VEV vanishes at $r\gg l_s$ and it
develops a nonzero VEV only near the $(p,q)$5-brane as 
$\chi\big|_{r\ll l_s}=-\frac{1}{g_s}\tan\theta$.
The low energy effective action contains
\begin{equation}
S_{{\rm D}3} = \int_0^{L_6}\int\left(
-\frac{1}{4g_4^2}F_{\mu\nu}F^{\mu\nu} - \frac{1}{2g_4^2}F_{\mu6}F^{\mu6} 
+ \frac{1}{4\pi} \epsilon^{\mu\nu\rho} A_\mu\p_\nu A_\rho \p_6\chi
 + \cdots\right)\d^3x \d x^6,
\label{eq:d3eff}
\end{equation}
where the integration range on $x^6$ is from $x^6=0$ to $x^6=L_6$.
Here $g_4^2=2\pi g_s$ is the $d=3+1$ gauge coupling constant on the
D3-branes. 
Ignoring all massive Kaluza-Klein (KK) modes of order $1/L_6$ or
smaller reduces the theory to the three dimensional Lagrangian with
gauge coupling constant $g^2=g_4^2/L_6$ as in eq.~\eqref{eq:gc}. 

The gauge transformation for $A_\mu$ reads
\beq
\delta S_{D3} &=& \int_0^{L_6} \int \dA_\mu \left(
-\frac{1}{g_4^2} \p_6 F^{\mu6} + \frac{1}{4\pi} \epsilon^{\mu\nu\rho}F_{\nu\rho}\p_6\chi
\right) \d^3x\d x^6 \nonumber\\
&=& \int\dA_\mu \left[- \frac{1}{g_4^2}F^{\mu6} + \frac{1}{4\pi}
\epsilon^{\mu\nu\rho}F_{\nu\rho}\chi\right]^{L_6}_0\d^3x.
\eeq
We must require this to vanish which gives the following boundary conditions
\beq
F_{\mu6} \big|_{x^6 = 0} = 0,\qquad
F_{\mu6} \big|_{x^6 = L_6} = \frac{g_4^2}{4\pi}\chi(L_6) \epsilon_{\mu\nu\rho} F^{\nu\rho} 
= -\frac{g_4^2}{4\pi g_s} \tan\theta\;\epsilon_{\mu\nu\rho} F^{\nu\rho}.
\eeq
The former is nothing but the boundary condition for the D3-brane
ending on the NS5-brane at $x^6=0$, while the latter can be understood
as an $\SL(2,\mathbb{Z})$ transformation of the boundary conditions by the
NS5-brane and D5-branes, namely the $(p,q)$5-brane. 
From eq.~\eqref{eq:d3eff}, we find that the CS coupling constant
induced on the D3-brane is
\beq
\kappa' = -\frac{1}{g_s} \tan\theta = -\frac{p}{q}.
\eeq
Thus, for the choice $(p,q)=(\kappa,1)$ the Chern-Simons coupling
constant is $\kappa$. 
The extension to multiple D3-branes, namely the non-Abelian theory, is 
straightforward. 

The hypermultiplets including squark fields, $(H,\tilde H^\dag)$,
which are $\NC\times \NF$ matrix-valued fields, correspond to
excitations of open strings between $\NC$ D3-branes and $\NF$
D5-branes. 
The Fayet-Iliopoulos (FI) parameter is related to the distance $L_7$
in the $x^7$ direction (we place the left NS5-brane at $x^7=0$ and the
right $(\kappa \to 0,1)$5-brane at $x^7=L_7$) between the left
NS5-brane and right $(\kappa,1)$5-brane 
\beq
v^2 = \frac{L_7}{4\pi^2 g_s l_s^2}.
\eeq
\begin{figure}[!htp]
\begin{center}
\includegraphics[width=\linewidth]{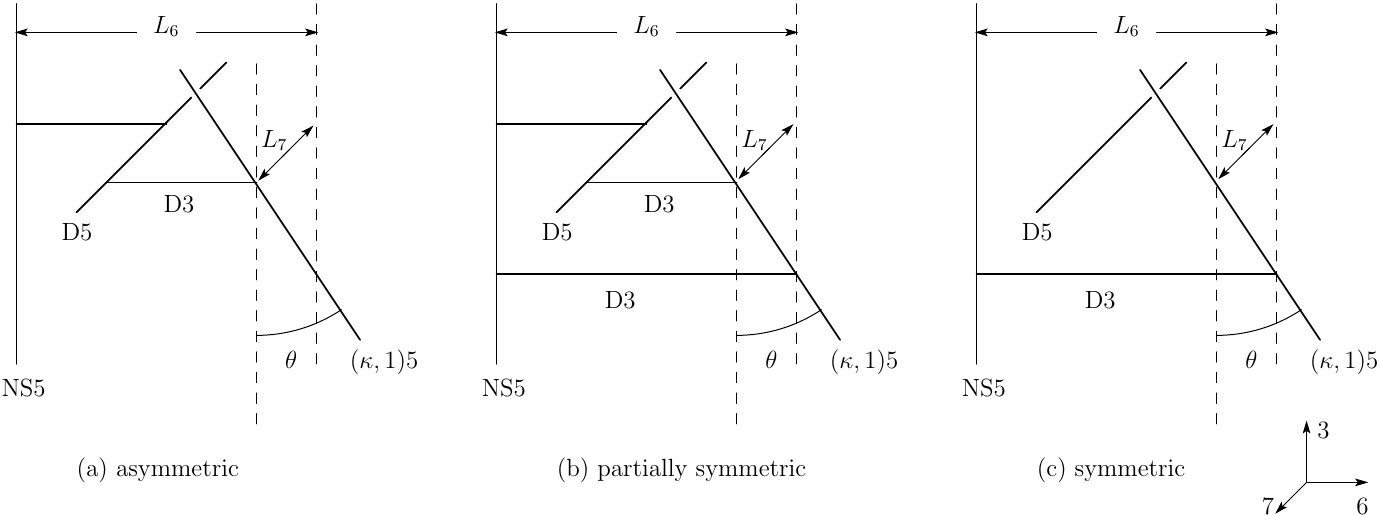}
\caption{{\footnotesize Three kinds of supersymmetric vacua can be
    geometrically understood as D-brane configurations in
    $(x^3,x^6,x^7)$-space. The $(\kappa,1)$5-brane is oriented in the
    $(x^3,x^7)$-space at angle $\theta$. The left NS5-brane is at
    $(x^6,x^7)=(0,0)$. The right $(\kappa,1)$5-brane is rotated to
    $(x^3,x^6,x^7)=(0,L_6,L_7)$}.
}
\label{fig:brane}
\end{center}
\end{figure}
The schematic picture is shown in fig.~\ref{fig:brane}.

The field theory limit can be taken by $g_s\to 0$ and $l_s\to 0$ with
the field theory parameters 
\beq
g^2 \sim \frac{g_s}{L_6}, \qquad
\kappa \sim \frac{g_s}{\theta}, \qquad
v^2 \sim \frac{L_7}{g_sl_s^2},
\eeq
being fixed.
The scalar potential in the low-energy effective theory is given by
\beq
V = \tr\left[
\frac{g^2}{2}\left(HH^\dagger - \tilde H^\dagger \tilde H - \frac{\kappa}{4\pi}X_3 - v^2{\bf 1}_N\right)^2
+ (HH^\dagger + \tilde H^\dagger \tilde H)X_3^2 
\right],
\eeq
where $X_3$ corresponds to the positions of $D3$-branes in the $x^3$
direction and is in the adjoint representation of $\U(\NC)$. 
If we identify $X_3$ with $\phi$ and set $\tilde H=0$, this
corresponds to the theory with the equal gauge coupling constants,
$e=g$, also considered in previous sections.

From fig.~\ref{fig:brane}, one can easily find three phases of the
classical supersymmetric vacua $V=0$:
\begin{itemize}
\item[(a)] asymmetric phase where $\U(\NC)$ is completely broken,
\item[(b)] partially symmetric phase where $\U(\NC)$ is broken to
  $\U(n)$ with $n\in[1,\NC-1]$,
\item[(c)] symmetric phase where the gauge group is fully unbroken.
\end{itemize}
One can also read off quantum effects by taking into account the s-rule.

Since we are interested in the topological solitons which arise due to
the spontaneously broken symmetry, we will consider only the
asymmetric phase in the rest of this section. 
The topological vortex in the Abelian case was identified with a
$(\kappa,1)$-string suspended by the $\NC=1$ D3-brane split on the
$\NF=1$ D5-brane \cite{Ohta:1999gj}. 
The axion background field again affects a D-string and it acquires
the electric charge proportional to $\chi$ by an analogue of Witten's
effect.
Indeed, the tension formula for the $(p,q)$-string in the constant
axion field background is given by \cite{Schwarz:1995dk} 
\beq
T_{(p,q)} = \frac{1}{2\pi l_s^2} \sqrt{\left(\frac{q}{g_s}\right)^2 + (p + q\chi)^2}.
\eeq
Thus the miminal tension is not that of a D-string but of a
$(\kappa,1)$-string \cite{Ohta:1999gj}. 
This is consistent with the field theory result that the mass of a
$(\kappa,1)$-string between the D3-branes is
\beq 
L_7 T_{(\kappa,1)} = \frac{L_7}{2\pi g_sl_s^2} = 2\pi v^2. 
\eeq
This is exactly the tension of a single topological vortex.

Now we will extend this result from the Abelian case to the
non-Abelian case.
There are now $\NC$ D3-branes and $\NF$ D5-branes.
The topological vortices are again identified with a
$(\kappa,1)$-string, and their electric charge and tension can be read
off of the string theory as done above.

\begin{figure}[!htp]
\begin{center}
\includegraphics[width=0.6\linewidth]{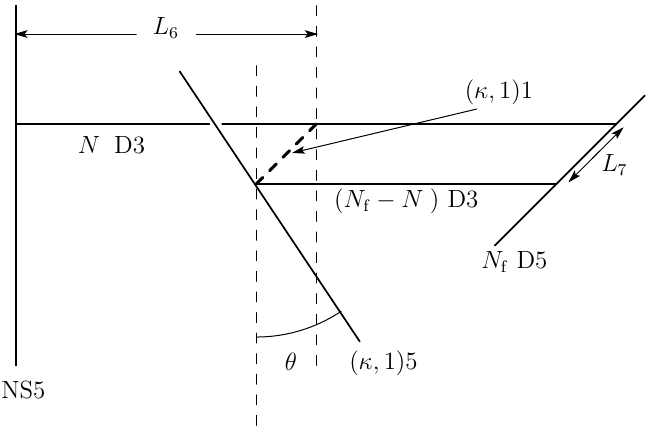}
\caption{\footnotesize The Hanany-Witten shift from
  fig.~\ref{fig:brane} (a). The $(\kappa,1)$-string (bold dashed line)
  corresponds to the topological non-Abelian local (for $\NC=\NF$) and
  semilocal (for $\NF>\NC$) vortices.
}
\label{fig:hw}
\end{center}
\end{figure}
The aim of the remainder of this section is to find the moduli space
of YMCSH vortices. 
From the D-brane perspective, the dimension of the moduli space is easily
counted as follows. 
Let us first move the D5-brane beyond the right $(\kappa,1)$5-brane in
the $x^6$ direction.  
Taking the Hanany-Witten effect into account, the brane configuration
results as shown in fig.~\ref{fig:hw}.

The topological vortex is identified with the $(\kappa,1)$-string
extending in $(x^1)$-space and suspended by the D3-branes as well as the
$(\kappa,1)$5-brane.
Note that if we send $\theta\to 0$ ($\kappa\to0$), 
fig.~\ref{fig:hw} reduces to the D-brane configuration of
ref.~\cite{Hanany:2003hp}, which is that describing non-Abelian
vortices in ${\cal N}=2$ supersymmetric Yang-Mills-Higgs theory
without the CS term. 
As in ref.~\cite{Hanany:2003hp}, the zeromodes of vortices are
identified with the zeromodes on the $(\kappa,1)$-string.
Open strings between $k$ $(\kappa,1)$-strings yield a $k\times k$
matrix field $Z$ and those between $k$ $(\kappa,1)$-strings and $\NC$
D3-branes ($\NF-\NC$ D3-branes) yield a $k \times \NC$ matrix field
$\psi$ ($(\NF-\NC) \times k$ matrix field $\tilde \psi$). 
The field $Z$ is in the adjoint representation of the $\U(k)$ gauge
symmetry while $\psi,\tilde\psi^\dag$ are in the fundamental
representation. 
These fields obey $k^2$ real D-term conditions.
Thus the complex dimension of the moduli space is
\beq
\dim_{\mathbb C}\mathcal{M}_{k,\NC,\NF} = k^2 + k\NC + (\NF-\NC) k - k^2 = k\NF.
\eeq
This moduli space is for the gauge group $G=\U(N)$, for which $n_0=\NC$
and hence $k N\NF/n_0=k\NF$.
The string theory result thus precisely coincides with the previous
results on the field theory side.

As we have seen, the D-brane picture serves as an easy way of counting
the dimensions of moduli space of topological solitons.
Nevertheless, it is also possible to construct a low-energy effective
theory on world volume of the soliton. 
For example, an effective theory for the non-Abelian vortices was made
using a Type IIB D-brane configuration in ref.~\cite{Hanany:2003hp}.

Collie and Tong \cite{Collie:2008mx} have derived a low energy
effective action on the world-volume of the non-Abelian CS vortex in
the field theory. 
In the rest of this section, we will attempt to derive it using the
brane configuration at hand. 
Especially, we are interested in the single vortex which carries
orientation zeromodes $\mathbb{C}P^{\NC-1}$ in addition to the
position zeromodes $\mathbb{C}$ when $\NC$ is equal to $\NF$.
The position zeromodes correspond to $Z$, which is simply a complex
field. The orientational zeromodes are given by $\psi$ and there are
no $\tilde\psi$ fields because $\NC=\NF$.

The low-energy effective theory of the CS vortex corresponds to the
low-energy effective theory on the $(\kappa,1)$-string shown in
fig.~\ref{fig:hw}. 
Let us first see that a CS term again appears in the low-energy
effective theory on a single $(\kappa,1)$-string.
To this end, we begin with the low-energy effective action on the
$(\kappa,1)$-string 
\beq
S_{(\kappa,1)1} = \int_0^{L_7}\int
\left(
\frac{1}{2g_2^2}f_{07}^2 + \frac{1}{4\pi}a_0\p_7\chi + \cdots
\right) \d t\d x^7,
\label{eq:effd1}
\eeq
where the integration range of $x^7$ is from 0 to $L_7$; the
$(\kappa,1)$-string ends on the NS5-brane at $x^7=0$ and on the
$(\kappa,1)$5-brane at $x^7=L_7$.
Here $a_\alpha$ ($\alpha=0,7$) are the Abelian gauge fields on the
$(\kappa,1)$-string and $g_2$ stands for the gauge coupling constant
$g_2^2=g_s/l_s^2$. 
Varying the action, we get
\beq
\delta S_{(\kappa,1)1} = \int\da_0 
\left[
- \frac{1}{g_2^2} f_{07} + \frac{1}{4\pi}\chi
\right]^{x^7=L_7}_{x^7=0}\d t.
\eeq
As before, the axion field acquires a nonzero VEV only in the region
close to the $(\kappa,1)$5-brane at $x^7=L_7$.
Then we can read off the boundary condition at $x^7=L_7$ as
\beq
f_{07} \big|_{x^7=L_7} = \frac{g_2^2}{4\pi} \chi(L_7) = - \frac{g_2^2}{4\pi} \kappa.
\eeq
This electric flux, proportional to $\kappa$, is responsible for the
electrically charged D-string, namely the $(\kappa,1)$-string.
From eq.~\eqref{eq:effd1}, the CS term in $d=1$ spacetime is found as
\beq
S_{\rm CS}^{d=1} = -\frac{\kappa}{4\pi} \int a = -\frac{\kappa}{4\pi} \int a_0\;\d t.
\eeq
We are now ready to construct the effective action for a single
non-Abelian CS vortex. 
Note that, for the $(\kappa,1)$-string, the distance $L_7$ is related
to the $\U(1)$ gauge coupling constant while $L_6$ is the FI term.
The $\U(1)$ gauge coupling constant $g_1$ in one dimension
is given by
\beq
g_1^2 = \frac{g_2^2}{L_7} = \frac{g_s}{l_s^2L_7}.
\eeq
This diverges in the decoupling limit $l_s\to 0$, since we are keeping
$v^{-2}\sim g_sl_s^2/L_7$ fixed.
The FI term in one dimension $v_1^2=L_6/g_s$ is related to the
3-dimensional gauge coupling constant via eq.~\eqref{eq:gc} as
$v_1^2=2\pi/g^2$ \cite{Hanany:2003hp}. 
Thus, the $\U(1)$ vector multiplet is infinitely heavy, which forces
the theory to remain on the Higgs branch, as in
ref.~\cite{Hanany:2003hp}.  
Thus the $\NC$ chiral multiplets $\psi_i$ (zeromodes of F-strings
between the D3 branes and the $(\kappa,1)$-string) must obey the
D-term constraint
\beq
\sum_{i=1}^{\NC} |\psi_i|^2 = \frac{2\pi}{g^2}.
\eeq
This, together with the $\U(1)$ gauge symmetry
$\psi_i\sim e^{\i\alpha}\psi_i$, means that $\psi_i$ are the fields
taking value in $\mathbb{C}P^{\NC-1}$.
In this way we obtain the $d=1$ effective Lagrangian 
\beq
L = \sum_{i=1}^{\NC} |D_0\psi_i|^2 - \frac{\kappa}{4\pi}a_0,
\eeq
with $D_0\psi_i=\p_0\psi_i+\i a_0\psi_i$.
Note that the gauge kinetic term vanishes because of the strong
coupling limit $g_1 \to \infty$.
This is the main result of this section and is in full agreement with
the result of ref.~\cite{Collie:2008mx}.

Note that we have shown, using string theory, that the CS couplings in
one dimension and in three dimensions are related and induced by the
$(\kappa,1)$5-brane.
This is also in agreement with the results of ref.~\cite{Collie:2008mx},
where field theory was used.

\section{Conclusion and discussion}\label{sec:conclusion}

In this paper, we have studied the zeromodes or moduli of the
YMCSH vortices. After setting up the model, we
discussed many limits of the theory, like the Yang-Mills-Higgs and
non-Abelian Chern-Simons-Higgs theories, found the mass spectrum of
the theory and wrote down the BPS equations in convenient form, valid
for any gauge group $G=(\U(1)\times G')/\mathbb{Z}_{n_0}$, with $G'$ a
semi-simple gauge group.
We then studied the zeromodes or moduli using a Callias-type index
theorem, the moduli matrix approach and a D-brane construction in Type
IIB string theory.
The main result of the paper is given by theorems \ref{thm:1} and
\ref{thm:2}, which prove that
the index counts the number of real zeromodes of an operator that is
constructed from the linear fluctuation equations, derived from the
BPS equations.
The index really counts the difference between the dimension of the kernel
of the operator that corresponds to the fluctuations and that of its
adjoint.
Theorems \ref{thm:1} and \ref{thm:2} prove that the adjoint operator does not have
zeromodes and hence the index is the total number of zeromodes in the
vortex solutions.
The number of zeromodes or the dimension of the moduli space,
coincides with that of Yang-Mills-Higgs theory and Chern-Simons-Higgs
theory.
The vanishing theorem in the case of Yang-Mills-Higgs theory is
rigorously proved (see refs.~\cite{Hanany:2003hp,Eto:2009bg}),
whereas in the Chern-Simons-Higgs case, the trick of writing the
adjoint operator squared as a positive definite sum of terms plus
boundary terms that vanish upon integration over the plane, does not
work and hence no rigorous proof has been made (see
ref.~\cite{Eto:2010mu}).
In YMCSH theory studied in this paper, the
trick also does not work, but a suitable gauge fixing choice made it
possible to simplify the fluctuation equation for the Higgs field enough so that
the proof could be established.
We then extended the moduli matrix formalism (see
e.g.~refs.~\cite{Eto:2005yh}) to include the extra adjoint field that
is possessed by YMCSH theory and the result is
consistent with the result of the index theorem calculation, that the
moduli are all contained in the moduli matrix, that comes from the
self-dual BPS equation.
That is, the adjoint field is fully determined by the other fields.
Finally, we extend the construction of ref.~\cite{Ohta:1999gj} of the
Abelian Chern-Simons-Higgs vortices in Type IIB string theory to the
non-Abelian case, where we confirm the dimension of the moduli space
(the number of zeromodes).
We then construct the low-energy effective Lagrangian on the vortices
in agreement with the result of Collie and Tong, who found the same
result using field theory \cite{Collie:2008mx}.

The obvious missing piece in the construction of non-Abelian vortices,
is the generalization from $G=\U(N)$ to
$G=(\U(1)\times G')/\mathbb{Z}_{n_0}$ (with $G'$ a semi-simple gauge
group) in Type IIB string theory using a D-brane setup.
This has not been possible yet, with or without the Chern-Simons term
and it is not known what should be done in string theory to make such
a generalization possible. 

In this paper, we have restricted to the case of $\NF=N$ flavors,
which simplifies the vacuum equations that for $\NF>N$ possess vacuum
moduli.
The $\NF>N$ case also always contains semilocal zeromodes.
The index theorem and the moduli matrix method studied in this paper,
are already valid for the $\NF>N$ case, but we have not considered the
vacuum moduli and the repercussions of that carefully here.
The brane construction was made for any $\NF\geq N$, but we have not
derived the low-energy effective theory on the vortex brane for the
$\NF>N$ case. We leave such extensions to future work.

Finally, we have written down the PDEs as master equations and given
examples in specific patches of certain moduli spaces.
The constructed system of equations account for exactly the right
number of moduli, found by the index calculation, by means of the
moduli matrix, barring that the master equation fields $\Omega$ and
$\Upsilon$ do not possess moduli. However, we have not proved this.
That is, we have not proved uniqueness of the master equations for
given fixed moduli matrix, fixing the $kN\NF/n_{0}$ known moduli.
We leave this problem as a direction of future work.

\subsection*{Acknowledgments}

S.~B.~G.~thanks the Outstanding Talent Program of Henan University for
partial support.
The work of S.~B.~G.~is supported by the National Natural Science
Foundation of China (Grants No.~11675223 and No.~12071111).
The work of M.~E.~is supported in part by JSPS Grant-in-Aid for
Scientific Research KAKENHI Grant No.~JP19K0383 and MEXT KAKENHI
Grant-in-Aid for Scientific Research on Innovative Areas ``Discrete
Geometric Analysis for Materials Design'' No.~JP17H06462 from the MEXT
of Japan.

\end{document}